\documentclass[11pt,a4paper]{article}
\pdfoutput=1
\usepackage{jcappub}
\usepackage[utf8]{inputenc}
\usepackage[english]{babel}
\usepackage{amsmath, amssymb}
\usepackage{graphicx}
\usepackage{hyperref}
\usepackage{caption}
\usepackage{float} 

\usepackage{xspace} 

\newcommand{\lcdm}{\ensuremath{\Lambda}CDM\xspace}

\newcommand{\dcdm}{\ensuremath{_{\mathrm{dcdm}}} }
\newcommand{\wdm}{\ensuremath{_{\mathrm{wdm}}} }
\newcommand{\dr}{\ensuremath{_{\mathrm{dr}}} }
\newcommand{\dx}{\mathrm{d}}

\subheader{\small
	\vspace*{-1.28cm}
	\begin{flushright}
		TUM-HEP-1420/22
	\end{flushright}
}

\title{Decaying Dark Matter and Lyman-\boldmath$\alpha$ forest constraints}

\author{Lea Fuß,}
\author{Mathias Garny}

\affiliation{Physik Department T31,\\
	James-Franck-Str.\ 1, Technische Universit\"at M\"unchen,\\
	D-85748 Garching, Germany}

\emailAdd{lea.fuss@tum.de}
\emailAdd{mathias.garny@tum.de}

\abstract{%
	Decaying Cold Dark Matter (DCDM) is a model that is currently under investigation regarding primarily the $S_8$ tension between cosmic microwave background (CMB) and certain large-scale structure measurements. The decay into one massive and one (or more) massless daughter particle(s) leads to a suppression of the power spectrum in the late universe that depends on the relative mass splitting $\epsilon=(1-m^2/M^2)/2$ between the mother and massive daughter particle as well as the lifetime $\tau$. In this work we investigate the impact of the BOSS DR14 one-dimensional Lyman-$\alpha$ forest flux power spectrum on the DCDM model using a conservative effective model approach to account for astrophysical uncertainties. Since the suppression of the power spectrum due to decay builds up at low redshift, we find that regions in parameter space that address the $S_8$ tension can be well compatible with the Lyman-$\alpha$ forest. Nevertheless, for values of the degeneracy parameter $\epsilon\sim 0.1-0.5\%$, for which the power suppression occurs within the scales probed by BOSS Lyman-$\alpha$ data, we find improved constraints compared to previous CMB and galaxy clustering analyses, obtaining $\tau\gtrsim 18$\,Gyrs for small mass splitting. Furthermore, our analysis of the BOSS Lyman-$\alpha$ flux power spectrum allows for values $\tau\sim 10^2$\,Gyrs, $\epsilon\sim 1\%$, that have been found to be preferred by a combination of Planck and galaxy clustering data with a KiDS prior on $S_8$, and we even find a marginal preference within this regime.
}

\begin{document}
	\maketitle
	
	\section{Introduction}
	\label{sec:intro}
	
	The standard model of cosmology known as \lcdm is a very successful model in explaining the large scale structure (LSS) of the universe. Cold dark matter (CDM) sits at the heart of this model, causing the typical hierarchical bottom-up structure formation we observe in the LSS by being non-relativistic during the clustering process. Despite the success of CDM, there are still unresolved issues that are hinting that there may be more~\cite{Abdalla:2022yfr}. This sparks interest in different cosmological models that are able to address these issues. Moreover, comparing the process of structure formation within extended cosmological models with LSS observations allows us to constrain fundamental properties of the two large unknowns, dark energy and dark matter, such as the equation-of-state, or the lifetime.
	
	One of the open questions is the so-called $\sigma_8$ tension, where $\sigma_8$ is a measure of the amplitude of matter fluctuations at a scale of $8$\,Mpc$/h$. More specifically, it is convenient to use the parameter $S_8 = \sigma_8 \sqrt{\Omega_m/0.3}$ that also includes the matter density parameter $\Omega_{m}$. The  tension arises between early universe cosmological data preferring larger values of $S_8$, and local, low redshift measurements tending towards lower values when interpreted within the $\Lambda$CDM model, with a typical significance of the order of $2-3\sigma$~\cite{Abdalla:2022yfr,DiValentino:2020vvd}. Measurements of the cosmic microwave background (CMB) temperature and polarization anisotropies by Planck yield $S_8 = 0.834 \pm 0.016$ \cite{Planck:2018vyg} which, in this respect, is in agreement with other CMB data like from ACT~\cite{ACT:2020gnv}. On the other hand, weak gravitational lensing surveys provide constraints via cosmic shear, e.g. $S_8 = 0.759^{+0.024}_{-0.021}$ from the Kilo-Degree Survey KiDS-1000~\cite{KiDS:2020suj} and $S_8=0.780^{+0.030}_{-0.033}$ from HSC~\cite{HSC:2018mrq}. The combination from shear and galaxy clustering from three-year data of DES yields $S_8=0.776^{+0.017}_{-0.017}$~\cite{DES:2021wwk} and a combination of shear, clustering and galaxy abundance from KiDS-1000 $S_8=0.773^{+0.028}_{-0.030}$~\cite{Dvornik:2022xap},
	while galaxy cluster counts from SPT-SZ report $S_8=0.766\pm0.025$~\cite{SPT:2018njh} and eROSITA results favour $S_8=0.791^{+0.028}_{-0.031}$~\cite{Chiu:2022qgb}. Individual measurements might not produce as large deviations, but show a trend  being significantly lower compared to CMB data. To account for this, and leaving aside the possibility of a statistical fluctuation, there either needs to be some unaccounted systematic error (see e.g.~\cite{Amon:2022azi}) or an alternative to $\Lambda$CDM featuring a suppression of the matter power spectrum in the $k\sim 0.1-1\,h/$Mpc regime.
	
	One model to achieve such a suppression is the Decaying Cold Dark Matter (DCDM) model. It is based on the hypothesis that dark matter can decay on cosmological time-scales into secondary dark sector particles. The decay products are assumed to be effectively stable on cosmological scales and, like the dark matter itself, sufficiently weakly coupled to visible matter to escape (in-)direct detection. However, the kinetic energy released in the decay process counteracts the growth of structures and leads to a suppression of the power spectrum. The model has been mainly investigated in two variants: a decay into massless secondaries that act as dark radiation (DR), or into a massless and a massive daughter. Depending on the mass splitting between mother and massive daughter particles, the latter acts as warm dark matter (WDM) being gradually produced in the decay process in the late universe. For both variants, the evolution in the early universe is identical to $\Lambda$CDM, thereby preserving its success in explaining the CMB and LSS on very large scales.
	Both models were studied regarding the $S_8$ and also the Hubble tension \cite{Schoneberg:2021qvd}, taking CMB, BAO and recently also galaxy clustering data into account.
	The decay into massless secondaries was investigated e.g.~in~\cite{Audren:2014bca,Enqvist:2015ara,Berezhiani:2015yta,Poulin:2016nat,Bringmann:2018jpr,Pandey:2019plg,Nygaard:2020sow,DES:2020mpv,Anchordoqui:2022gmw,Simon:2022ftd,Alvi:2022aam}, allowing also the possibility that only a fraction $f$ of the dark matter decays. It was found that the decay into purely DR will most likely not be able to solve cosmological tensions and requires minimum lifetimes of around $\sim200$\,Gyrs for $f=1$. The latest works~\cite{Simon:2022ftd,Alvi:2022aam} for this model confirm this even more while also providing tight constraints. For lifetimes shorter than the age of the universe,~\cite{Simon:2022ftd} finds $f<2.16$\% and for $f\to 1$ a lower bound of $\tau > 250$\,Gyrs~\cite{Simon:2022ftd,Alvi:2022aam}.
	As a further variant, also the decay of warm dark matter mother particles into massless dark radiation has been considered~\cite{Blinov:2020uvz}, but similar to the decay of CDM into massless daughters, this setup was found to neither solve the $H_0$ nor $S_8$ tensions~\cite{Holm:2022eqq}.
	
	The model with a decay of CDM into WDM and DR, on which we mainly focus in this work, is apart from the lifetime $\tau$ described by the mass splitting parameter
	\begin{equation}
		\epsilon = \frac{1}{2} \left( 1 - \frac{m^2}{M^2} \right)\,,
		\label{eq:epsilon}
	\end{equation}
	involving the mass of the mother ($M$) and the massive daughter ($m$) particle.
	The results regarding the Hubble tension are similar compared to the massless case, implying that it is probably not to be resolved with DCDM~\cite{Haridasu:2020xaa,Clark:2020miy,Davari:2022uwd} (see also~\cite{Blackadder:2014wpa,Vattis:2019efj} for earlier work). The situation for the $S_8$ tension is however not so clear. While~\cite{Davari:2022uwd} suggests that this tension can also not be addressed,~\cite{Abellan:2020pmw}, which includes an improved treatment of perturbations, finds that it actually can be lessened for $\tau \sim 55$\,Gyrs and $\epsilon\sim 0.7$\% based on Planck CMB, BAO, RSD and SN Ia data. In two follow-up works a newly developed code for much faster computation of the DCDM power spectra is used. This allows for a more in depth analysis, like in~\cite{FrancoAbellan:2021sxk}, where a mild preference for DCDM is found depending on the priors for $S_8$. The latest work~\cite{Simon:2022ftd} also includes full-shape information from BOSS DR12 galaxy clustering, and finds that DCDM can ease the $S_8$ tension even though it is not performing significantly better than \lcdm{} when disregarding KiDS data. However, when including KiDS, DCDM is preferred, and the best-fit model occurs for a lifetime of $\tau \sim 120$\,Gyrs and $\epsilon\sim 1.2$\%.
	
	Another possibility to study DCDM is via galaxy and halo properties with more regards towards the small scale issues (see e.g.~\cite{Bullock:2017xww}) like the cusp-core problem of DM halos~\cite{deBlok:2009sp}. For DCDM with a massive and a massless daughter particle,~\cite{Peter:2010sz} connects the model to the observed population of Milky Way satellites.
	Combining numerical and semi-analytic methods, they find constraints at $\tau \gtrsim 30$\,Gyrs for $20 \lesssim v_k \lesssim 200$km/s. Here $v_k$ is the so-called kick-velocity  which is transferred to the daughter particles during the decay, being related to $\epsilon$ via $v_{k} \sim \epsilon c$ for $\epsilon\ll 0.5$. The analysis~\cite{Mau:2022sbf} builds up on this work and uses Milky Way satellite galaxies observed by DES, excluding $\tau < 18$\,Gyrs for $v_k=20$\,km/s. This probe is extremely sensitive to the low $\epsilon$ regime which still affects the halo distribution and substructure due to the low virial velocities in dwarf galaxies.
	
	Even when not considering cosmological tensions, it is still interesting to constrain fundamental properties of dark matter like its lifetime via different complementary probes, regarding how little we know about the actual particle nature, and the fact that very few particles are naturally stable~\cite{Audren:2014bca}. Therefore, the degree to which DCDM is compatible with various cosmological and astrophysical observations is worth studying.
	
	In this work we confront DCDM with measurements of the one-dimensional Lyman-$\alpha$ forest flux power spectrum, using data from BOSS DR14~\cite{Chabanier:2018rga}. The Lyman-$\alpha$ forest is an important probe for dark matter models that lead to a modification of the power spectrum on scales $k\gtrsim 1h/$Mpc, which is typically the case for models addressing the $S_8$ tension. A pecularity of DCDM is that the suppression of the power spectrum occurs at late redshifts, such that it is different for weak lensing,  galaxy clustering and cluster number count observations that are sensitive mainly to $z\lesssim 1$ as compared to Lyman-$\alpha$ measurements at $z\sim 2-4$. Therefore, one expects that a larger amount of power suppression is possible at low redshift as compared to models where the power suppression is imprinted already in the early universe, making DCDM a promising model in view of the $S_8$ tension and Lyman-$\alpha$ constraints.
	
	The main challenge in any Lyman-$\alpha$ forest analysis is the extraction of the actual matter fluctuations from the measured flux power spectrum, requiring a description of the complex intergalactic medium (IGM). In this work, we make use of an effective model that was already used to analyse BOSS data and extensively validated against hydrodynamical simulations for a variety of dark matter models as well as massive neutrino cosmologies in the past \cite{Garny:2020rom,Garny:2018byk}. It contains a number of free parameters that account for the IGM behavior as well as uncertainties from strongly non-linear scales entering via the line-of-sight projection, while taking advantage of the increased reach of a perturbative treatment of the underlying three-dimensional matter distribution at the relevant redshifts $z\sim 3$. This allows us to determine robust constraints on the DCDM parameters from the Lyman-$\alpha$ forest on the relatively large scales measured with a high precision by BOSS.
	
	The possibility to address the $S_8$ tension with DCDM raises the question about an embedding of this scenario in a more complete particle physics framework. We make a first step in this direction by exploring the generalization from two- to three-body decays, that generically occur in models where the involved particle species are fermions. A small mass splitting $\epsilon$ can be realized naturally by a pseudo-Dirac fermion pair in that setup.
	
	The structure of this work is as follows: In Sec.\,\ref{sec:dcdm}, we give an overview of the formalism of DCDM, the basic background dynamics and the generated power spectrum. Then, in Sec.\,\ref{sec:Lya}, we review the data set used in this work as well as the effective model and its input and free parameters. Afterwards, in Sec.\,\ref{sec:results} we present our results within the DCDM parameter space of lifetime and mass splitting. We also set them in context with earlier works with emphasis on the $S_8$ tension. In Sec.\,\ref{sec:3bodydecay} we comment on an extension from two- to three-body decays. Finally, we conclude in Sec.\,\ref{sec:conclusion}.
	
	\section{Decaying Cold Dark Matter}
	\label{sec:dcdm}
	
	\subsection{Formalism}
	\label{sec:formalism}
	
	The DCDM model we study comprises collisionless cold dark matter particles that are unstable and  decay into two components,
	\begin{equation}
		\text{DCDM} \ \rightarrow \ \text{WDM} \ + \ \text{DR}\,.
	\end{equation}
	One is a massive daughter acting as a warm dark matter (WDM) component whereas the other is massless dark radiation (DR, see Sec.\,\ref{sec:3bodydecay} for an extension to three-body decays).
	This model can be described by introducing two new parameters $\Gamma$ and $\epsilon$.
	The first one, $\Gamma$, is the decay width of the CDM mother particle which we usually replace by the decay time $\tau=\Gamma^{-1}$. It determines when the decay sets in. The second parameter $\epsilon$ defined in~\eqref{eq:epsilon} is related to the mass splitting between the mother and massive daughter particle, and characterizes the ratio of energy transformed into DR and WDM. The $\epsilon$ parameter is also related to the amount of energy that is transformed from rest mass into kinetic energy, and only depends on the mass ratio of mother and daughter and not on the absolute mass values. In the case of $m \to 0$ corresponding to $\epsilon \to 0.5$, only dark radiation is produced by a decay into two massless daughters.  In the opposite case of $m \to M$ corresponding to $\epsilon \to 0$, the daughter particle has almost the same mass as the mother particle and therefore the energy transferred to DR vanishes. Effectively, the decay becomes irrelevant for $\epsilon=0$, independently of the lifetime. 
	Therefore, in both the limits of $\tau \to \infty$ as well as $\epsilon \to 0$ one recovers \lcdm.
	
	\begin{figure}[t]
		\centering
		\includegraphics[height=0.3\textheight]{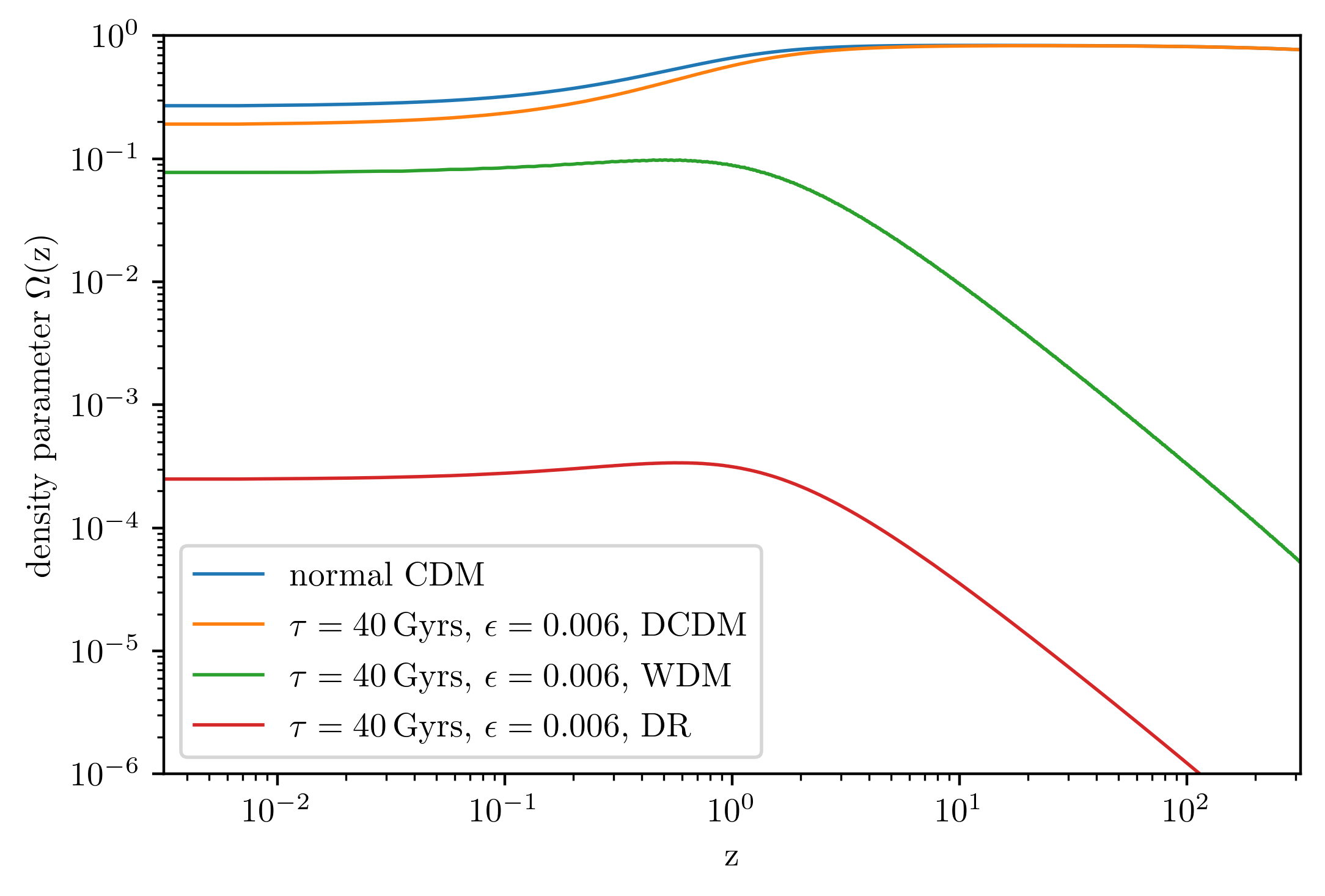}
		\caption{Background evolution of the DCDM (orange), WDM (green) and DR (red) density parameter for $\tau=40$\,Gyrs and $\epsilon=0.006$ in comparison to a conventional CDM scenario without any decay (blue). For large redshifts (corresponding to $t\ll\tau$) the decay is irrelevant, while subsequently the DCDM density drops below CDM, and WDM as well as DR are produced. The plateau at low redshift is due to the logarithmic $z$-axis.}
		\label{fig:DCDM_bg}
	\end{figure}
	
	Following~\cite{FrancoAbellan:2021sxk}, we work in synchronous gauge which is comoving with the mother particle. The physical energy-momentum four vectors then take the form $P\dcdm = (M, 0)$, $P\wdm = (\sqrt{m^2 + p^2}, \vec{p})$ and $P\dr = (p, -\vec{p})$. Therefore, the
	physical momentum $p\equiv |\vec p|$ of the daughter particles is fixed by energy and momentum conservation to
	\begin{equation}
		p_\text{2-body} = \frac{1}{2 M}(M^2 - m^2) = M \epsilon.
		\label{eq:pmax}
	\end{equation}
	The density parameter of DCDM today at $t_0$ is given by an initial density times an exponential factor describing the decay
	\begin{equation}
		\Omega^0\dcdm = \Omega\dcdm^\text{ini} e^{-\Gamma t_0}.
		\label{eq:Omega0dcdm}
	\end{equation}
	We can characterize the DCDM energy content by $\omega\dcdm^\text{ini}=\Omega\dcdm^\text{ini} h^2$ which is the density of DCDM today if no decay had taken place.
	At an arbitrary time the density is given by
	\begin{equation}
		\rho\dcdm = \rho_{\mathrm{crit, }0} \Omega\dcdm^{\mathrm{ini}} e^{-\Gamma t} a^{-3} = M\bar{N}_\text{dcdm} a^{-3}\,,
		\label{eq:omegadcdm}
	\end{equation}
	where $a^{-3}$ gives the additional expansion factor. It can alternatively be written as the time dependent number density $\bar{N}\dcdm$ times the rest energy of the mother particle.
	Next, we consider the Boltzmann equations relating the total time derivative of the phase-space distribution function to the collision term governed by the decay. At the homogeneous background level, they are given by
	\begin{flalign}
		\dot{\bar{f}}\dcdm(q,\tau) &= -a\Gamma\bar{f}\dcdm(q,\tau) \nonumber \quad \mathrm{and} \\
		\dot{\bar{f}}\wdm(q,\tau) = \dot{\bar{f}}\dr(q,\tau) &= \frac{a\Gamma\bar{N}\dcdm}{4\pi q^2} \delta(q- ap_\text{2-body})\,,
		\label{eq:BE}
	\end{flalign}
	where $q=ap$ is the comoving momentum.
	The collision term for DCDM only depends on $\Gamma$ due to the exponential decay. The scale factor $a$ arises from switching to conformal time with $\mathrm{d}\tau = \mathrm{d}t/a$, and we use the notation where a dot denotes $d/d\tau$. For WDM and DR, the collision term has the opposite sign and is proportional to the number density of DCDM. The factor $1/4 \pi q^2$ takes into account the spherical symmetry and the comoving momentum $q$ is fixed by the delta function to the momentum transferred to the daughter particles $ap_\text{2-body}=\epsilon aM$ due to two-body kinematics.
	The phase-space-distribution is related to the mean energy density $\bar{\rho}$ and pressure $\bar{P}$ by the integrals
	\begin{flalign}
		\bar{\rho} &= \frac{1}{a^4}\int_{0}^{\infty} \mathrm{d}q 4\pi q^2 {\cal E} \bar{f} \quad \mathrm{and} \nonumber\\
		\bar{P} &= \frac{1}{3a^4} \int_{0}^{\infty} \mathrm{d}q 4\pi q^2 \frac{q^2}{\cal E} \bar{f}.
		\label{eq:density_pressure}
	\end{flalign}
	Here ${\cal E} \equiv \sqrt{m^2a^2+q^2}$ is the comoving energy which is related to the physical energy by ${\cal E}=aE$.
	Making use of these definitions we can transform the Boltzmann equations to
	\begin{flalign}
		\dot{\bar{\rho}}\dcdm &= -3\mathcal{H} \bar{\rho}\dcdm - a\Gamma\bar{\rho}\dcdm\,, \nonumber\\
		\dot{\bar{\rho}}\dr &= -4\mathcal{H} \bar{\rho}\dr + \epsilon a\Gamma\bar{\rho}\dcdm\,, \nonumber\\
		\dot{\bar{\rho}}\wdm &= -3(1+\omega)\mathcal{H} \bar{\rho}\wdm + (1-\epsilon) a\Gamma\bar{\rho}\dcdm\,,
		\label{eq:densityDE}
	\end{flalign}
	with ${\mathcal H}=aH$, Hubble rate $H$, and  equation-of-state parameter $\omega=\bar{P}\wdm/\bar{\rho}\wdm$ for WDM  (and assuming an equation-of-state parameter equal to zero for DCDM as usual, which amounts to neglecting its velocity dispersion). Here, we see the impact of DCDM in the second term on the right-hand side, reducing or adding to the energy densities depending on $\Gamma$ and $\epsilon$. 
	
	\begin{figure}[t]
		\centering
		\includegraphics[height=0.3\textheight]{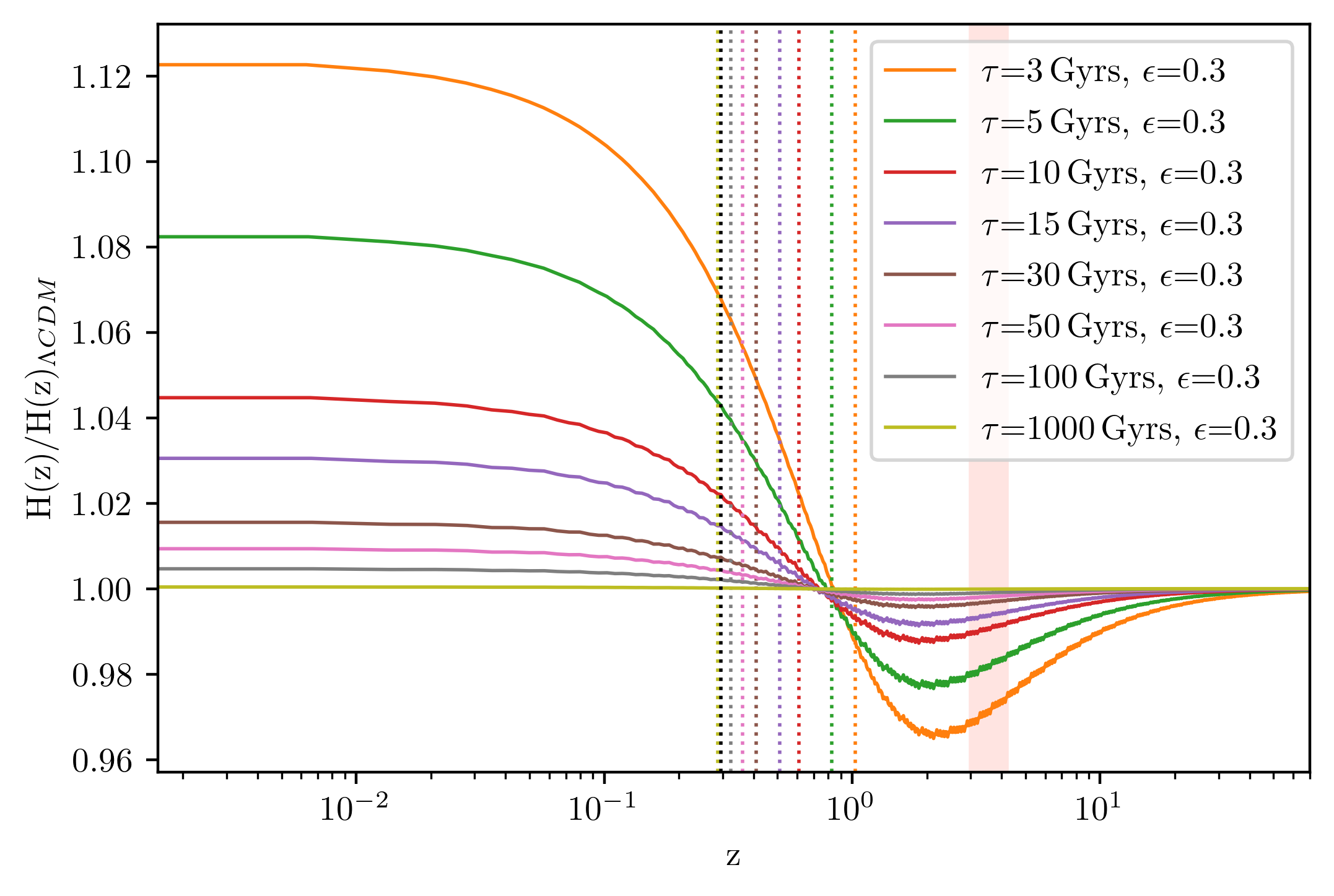}
		\caption{Hubble rate $H(z)$ depending on redshift normalized to \lcdm for various lifetimes $\tau$ and a large value $\epsilon=0.3$ of the mass splitting. The dotted colored lines correspond to the equality of matter and dark energy and the pink region to the redshifts spanned by the Lyman-$\alpha$ data used in this work. During matter domination the ratio decreases due to a shift from matter to radiation which lowers the energy density. To account for this change while keeping a fixed angular diameter distance to the last scattering surface the ratio has to increase for low redshifts. We note that for small values $\epsilon\ll 0.5$ relevant for the $S_8$ tension the modification of $H(z)$ is negligibly small as compared to the case $\epsilon=0.3$ shown here for illustration. The main change occurs at the perturbation level in that case, see Fig.\,\ref{fig:DCDM_Pk_e}.}
		\label{fig:DCDM_H}
	\end{figure}
	
	The resulting evolution of the energy density parameters is shown in Fig.\,\ref{fig:DCDM_bg} for $\tau=40$\,Gyrs and $\epsilon=0.006$, versus redshift $z$. Additionally, the cold dark matter density for \lcdm is shown in blue. For large redshifts, DCDM clearly converges towards \lcdm as expected. Only for redshifts long after recombination does the decay cause a drop in the DCDM density while WDM and DR are produced in return. Note that the densities approach a plateau for low redshift due to the logarithmic scale and since $z\to 0$ corresponds to the limit $t\to t_0$. The lifetime $\tau$ determines when the deviation from \lcdm sets in, while $\epsilon$ controls the relative size of the WDM and DR densities.
	
	The modified background quantities also result in a different Hubble expansion rate as compared to $\Lambda$CDM, with the difference being more pronounced the larger $\epsilon$ (corresponding to a larger fraction of DR) and the smaller $\tau$. The evolution of $H(z)/H(z)_{\Lambda\text{CDM}}$ for a large value $\epsilon=0.3$ is shown in Fig.\,\ref{fig:DCDM_H} for various lifetimes, including rather extreme values for illustration. Here we adjusted the dark energy density within DCDM such that the angle under which the first CMB peak appears is identical to the reference $\Lambda$CDM model in all cases, i.e.~all models feature an identical angular diameter distance at times $t\ll \tau$. The evolution at high redshift is identical to $\Lambda$CDM. Once the decay starts to set in, some amount of matter is replaced by radiation, which redshifts faster and subsequently contributes less to the energy content. This leads to a decrease of $H(z)/H(z)_{\Lambda\text{CDM}}$, lasting until dark energy becomes relevant (marked by the dotted vertical lines). To keep the angular diameter distance to the last scattering surface constant in all models, the dark energy content needs to be larger for shorter $\tau$. This leads to an increase of $H(z)/H(z)_{\Lambda\text{CDM}}$ at low redshift that over-compensates the earlier suppression. While this increase has been considered as a possibility to address the $H_0$ tension in the past, the increased dark energy content is not consistent with a combination of CMB, BAO and SN Ia data, as discussed previously. For $\epsilon\ll 0.5$, the regime that is mostly relevant for addressing the $S_8$ tension, the modification of $H(z)$ is almost negligible, since most of the energy density is transferred to the WDM component.
	The increase in $\Omega_\Lambda$ additionally leads to an increased late Integrated-Sachs-Wolfe effect in the CMB anisotropy spectrum. Again, for realistic values of $\tau$ and $\epsilon$ this effect is very small, similar to the effect on the Hubble expansion rate. A deviation from \lcdm{} occurs instead at the perturbation level, leading to a suppression in the power spectrum at late times.
	
	\subsection{The DCDM power spectrum}
	\label{sec:powerspectrum}
	
	\begin{figure}[t]
		\centering
		\includegraphics[height=0.25\textheight]{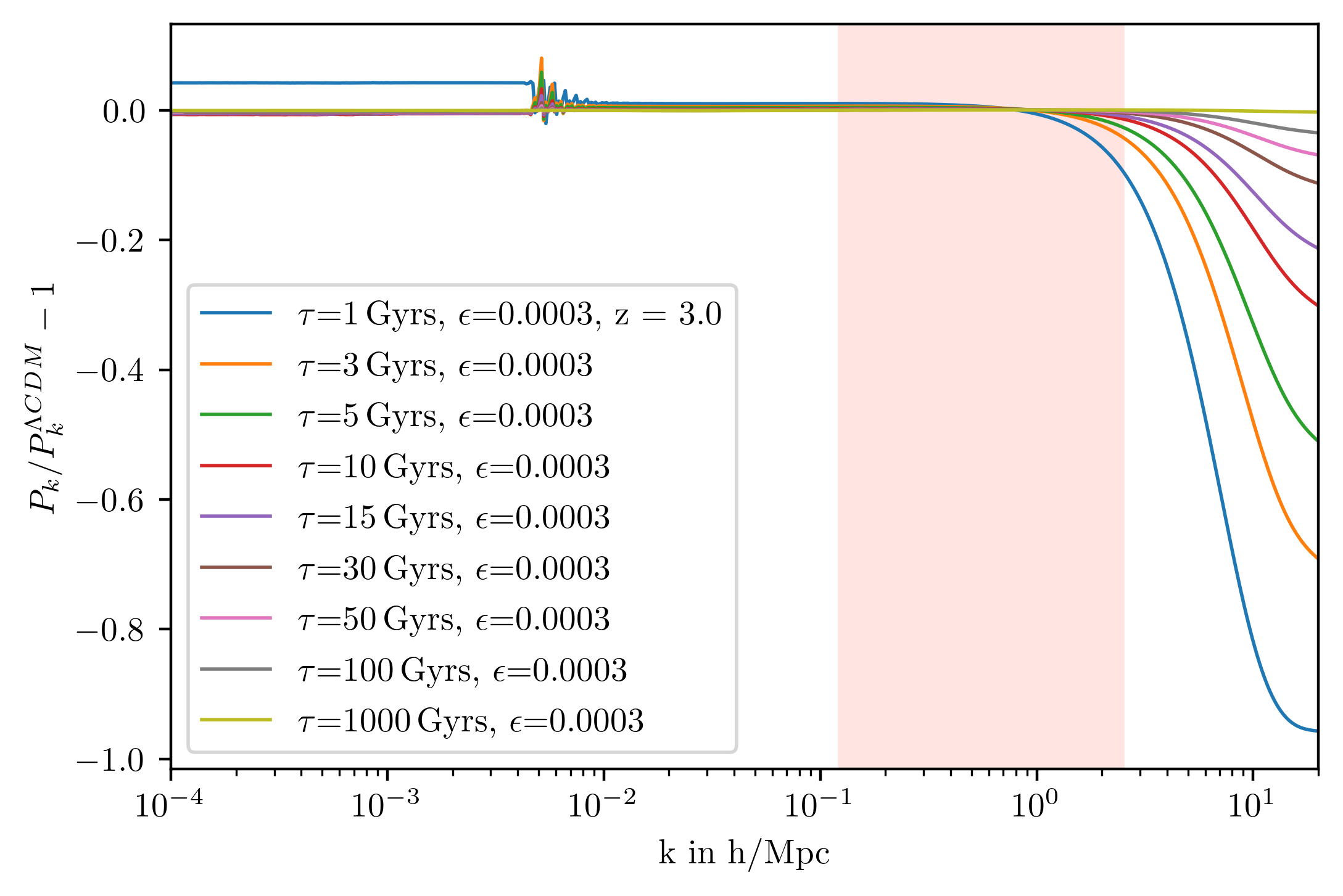}
		\includegraphics[height=0.25\textheight]{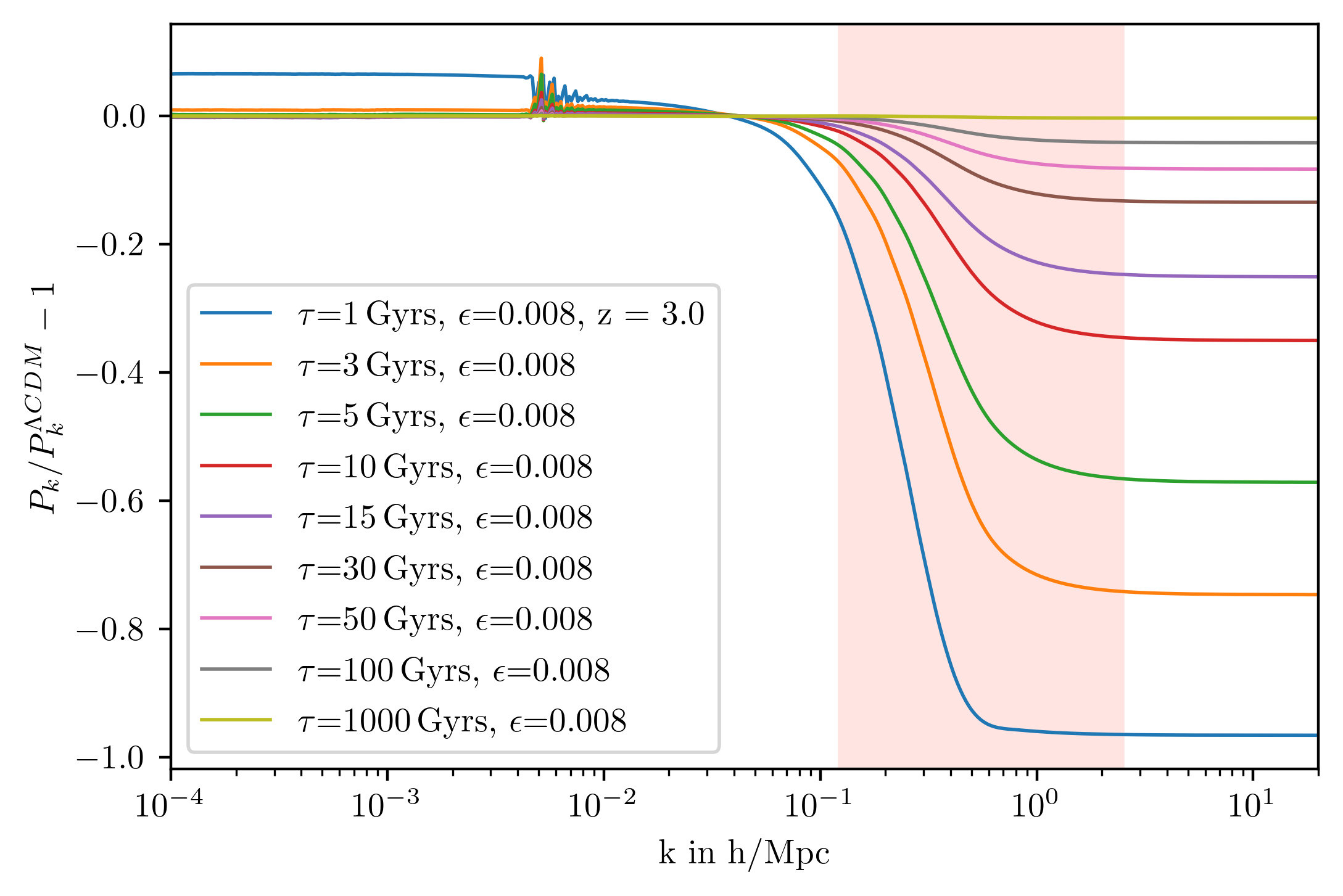}
		\includegraphics[height=0.25\textheight]{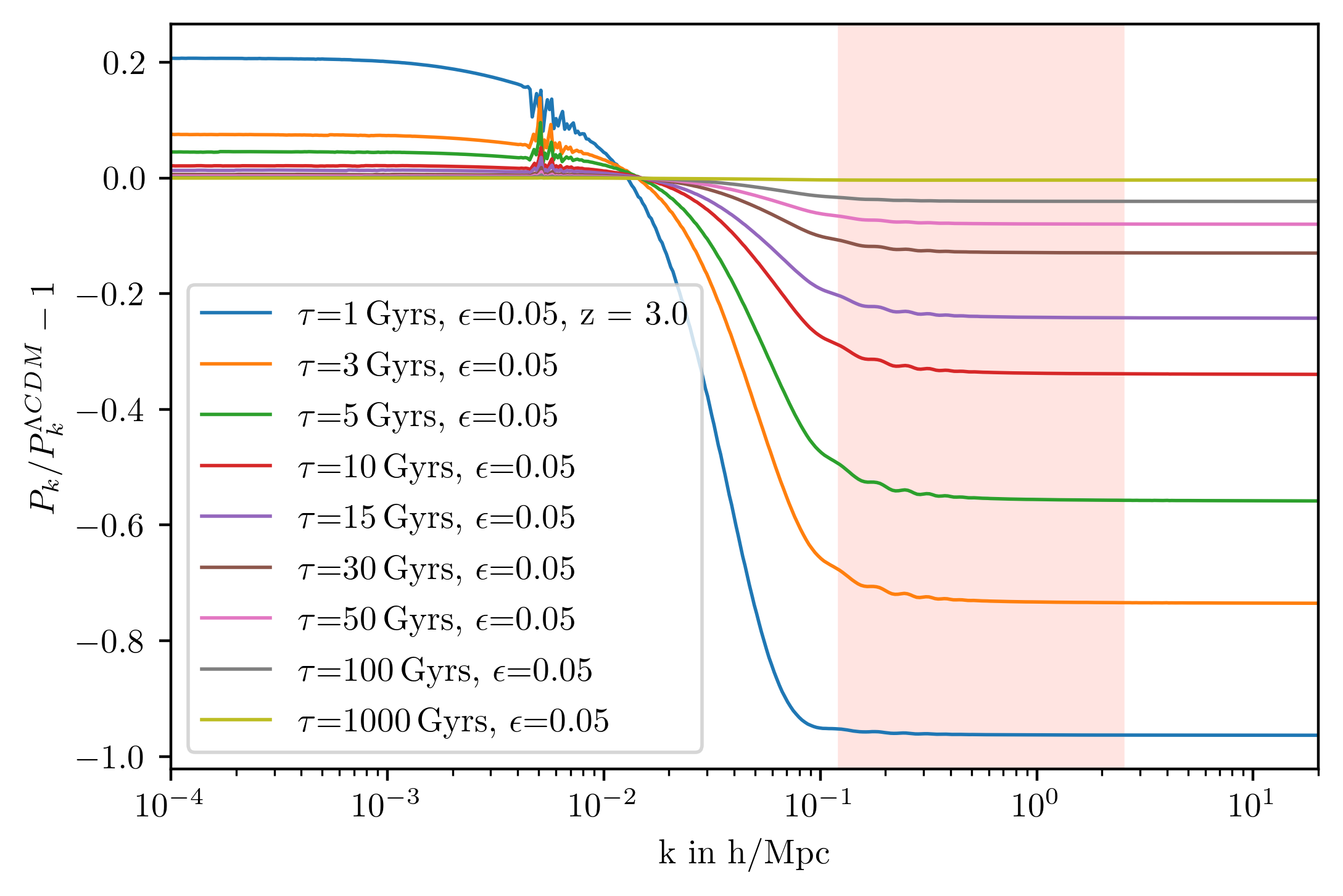}
		\caption{DCDM power spectrum normalized to $\Lambda$CDM at $z=3$ and for $\epsilon=0.0003, 0.008, 0.05$ from top to bottom, respectively. Each panel shows various lifetimes $\tau$. The pink region indicates the range of BOSS Lyman-$\alpha$ data. Since the $\epsilon$ parameter determines the kinetic energy of WDM produced in the decay, it controls the onset of suppression in the power spectrum. Larger $\epsilon$ corresponds to more kinetic energy, shifting the suppression scale to the left. The decay time is in turn responsible for the steepness of the suppression since it controls the amount of WDM at a given time.}
		\label{fig:DCDM_Pk_e}
	\end{figure}
	
	To obtain the total matter power spectrum we solve linear perturbation equations for the mother and daughter particles involved in the decay, coupled to metric perturbations in the usual way~\cite{Ma:1995ey}. For the decaying cold dark matter species it is sufficient to solve the standard continuity equation for its density contrast $\delta_\text{dcdm}$, given the synchronous gauge choice adopted here. For the daughter particles, the perturbation equations are similar in form to the usual perturbation equations for massless or massive neutrinos, for the dark radiation and warm dark matter component, respectively.
	In particular, they take the form of a coupled hierarchy for the multipole moments of the perturbed phase-space distribution function. The only difference to the case of neutrinos occurs for the equation of the monopole, that contains an additional source term on the right-hand side due to the decay, given by $-a\Gamma \bar f_\text{ dcdm} \delta_\text{ dcdm}$~\cite{FrancoAbellan:2021sxk}. Additionally, the homogeneous parts of the distribution functions are time-dependent, as described by~\eqref{eq:BE}.
	
	In addition to the coupled set of equations for the multipole moments, we consider a fluid approximation for the warm dark matter component as proposed in~\cite{FrancoAbellan:2021sxk}. It amounts to a coupled set of equations for the density contrast and velocity divergence of the massive daughter particle, being the usual continuity and Euler equations. The Euler equation is complemented with an effective pressure term, with sound velocity that depends on the wavenumber $k$. We adopt the choice given in equation (38) in~\cite{FrancoAbellan:2021sxk}, that was found to reproduce the result of the full coupled hierarchy to a high accuracy, while being computationally much less expensive.
	
	To solve the perturbation equations we use the modified CLASS code presented in~\cite{FrancoAbellan:2021sxk} by Abellán, Murgia and Poulin.\footnote{ \href{"https://github.com/PoulinV/class_decays"}{https://github.com/PoulinV/class\_decays}.} It includes an implementation of the DCDM model in CLASS based on the coupled hierarchy of multipole moments as well as the fluid approximation mentioned above. 
	We checked the agreement of both computational methods, and the dependence on precision parameters like momentum bins or the largest considered multipole moment in case of the computation based on the full hierarchy. We found that in the latter case numerical random oscillations, that may occur on smaller scales depending on the precision settings, are smoothed out for the fluid approximation. Since we are interested in the amount of power suppression in that regime and such oscillations might lead to unphysical artifacts for finite precision settings we found the much faster fluid approximation to be more suitable for our purpose. To generate the linear power spectrum for the \lcdm model we use the standard \href{"https://lesgourg.github.io/class_public/class.html"}{CLASS} code \cite{CLASS:2011,Lesgourgues:2011re}. 
	
	\begin{figure}[t]
		\centering
		\includegraphics[height=0.28\textheight]{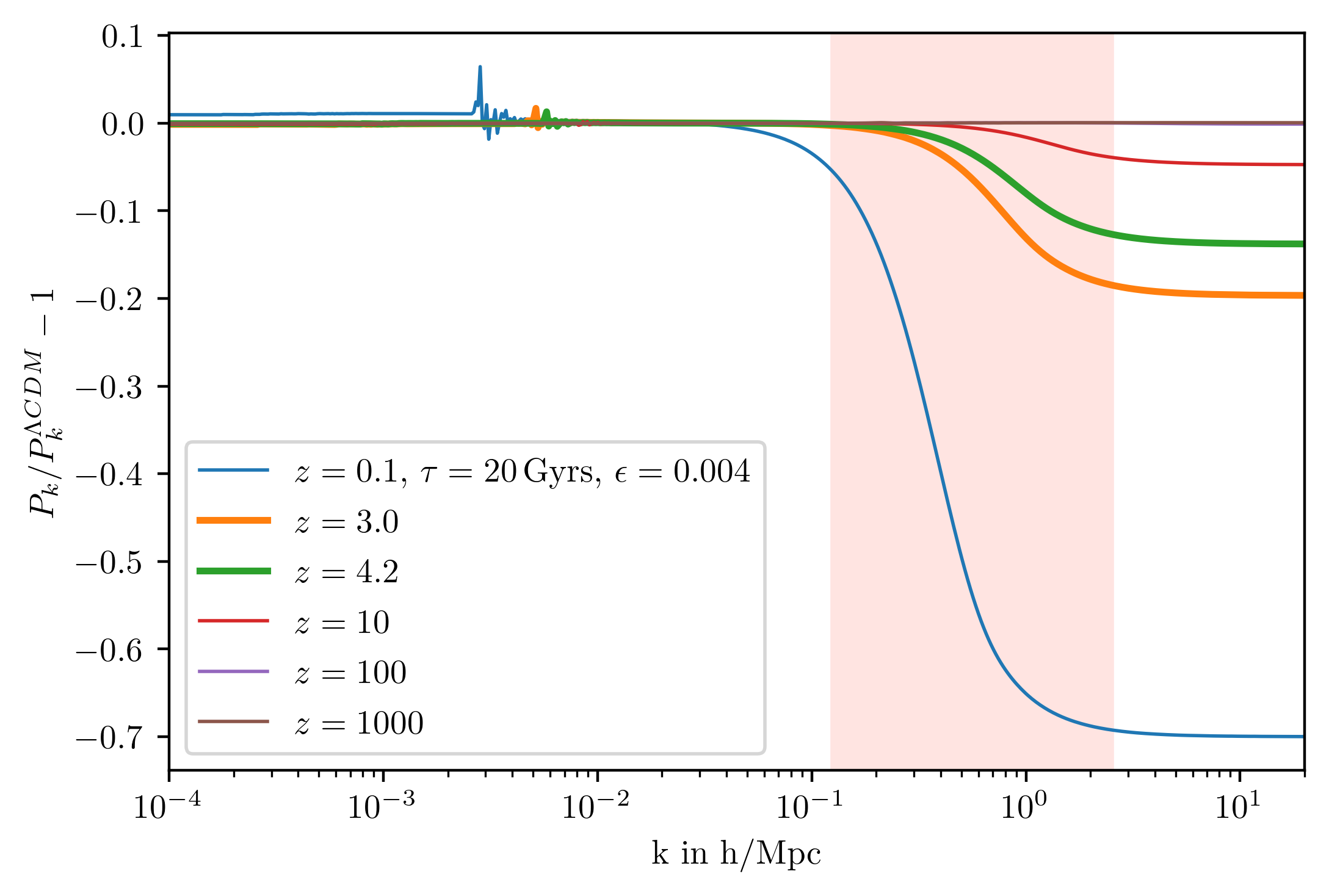}
		\caption{The DCDM power spectrum at $\tau=20$\,Gyrs and $\epsilon=0.004$ at $z=0.1, 3.0, 4.2, 10, 100, 1000$ relative to \lcdm at the respective redshifts. The relative suppression increases with time due to the ongoing decay. The thicker lines indicate the redshift range between $z=3.0$ and $z=4.2$ used in our Lyman-$\alpha$ analysis.}
		\label{fig:DCDM_Pk_zevolution}
	\end{figure}
	
	\begin{figure}[ht]
		\centering
		\includegraphics[height=0.28\textheight]{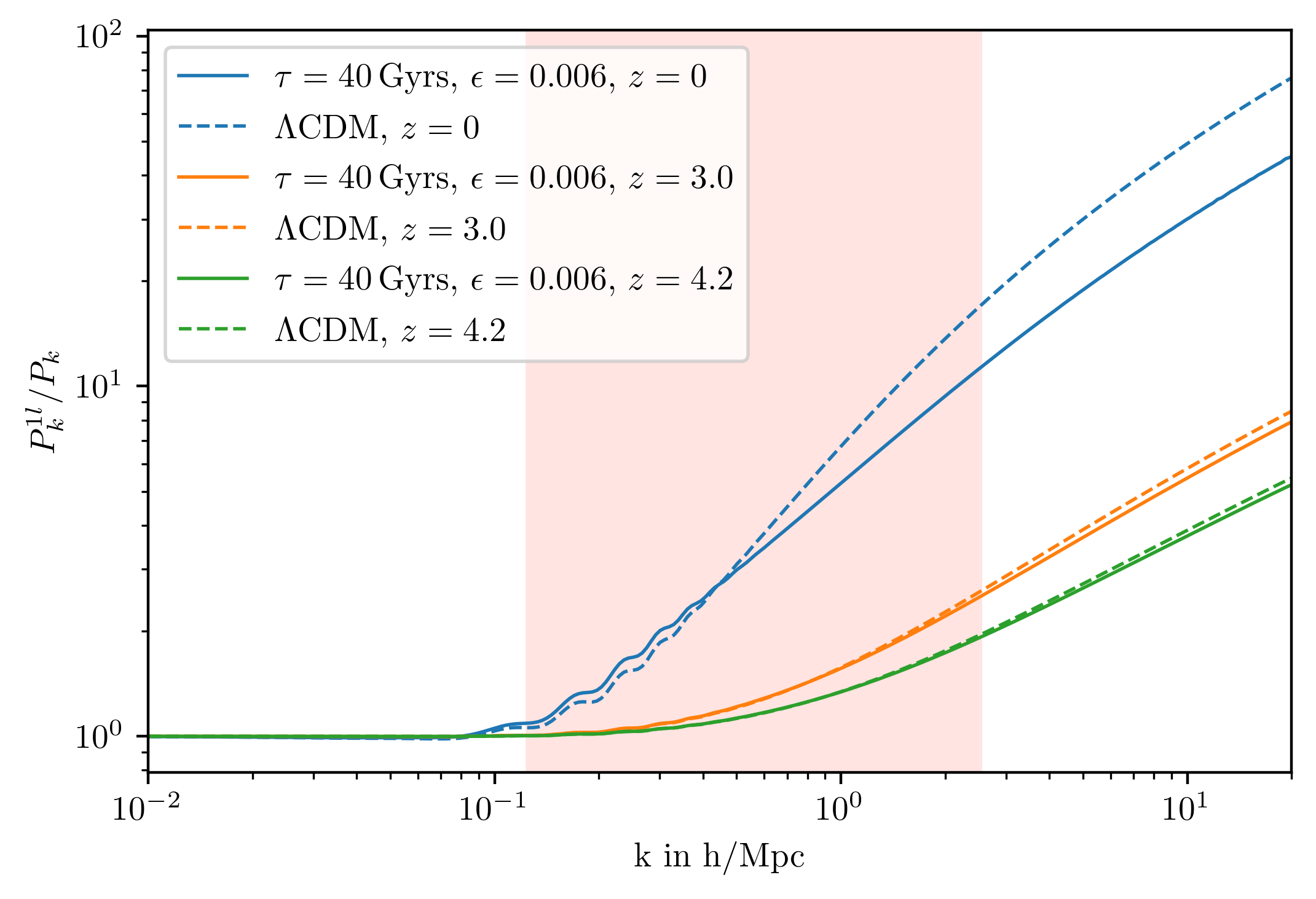}
		\caption{One-loop density power spectrum $P_{\delta\delta}/P_\mathrm{linear}$ normalized to the linear power spectrum for $\tau=40$\,Gyrs and $\epsilon=0.006$ (solid) and \lcdm (dashed) at $z=0, 3.0, 4.2$, respectively.}
		\label{fig:DCDM_Pk_znl}
	\end{figure}
	
	The resulting linear power spectra for several parameter combinations of lifetime $\tau$ and mass degeneracy $\epsilon$ are shown in Fig.\,\ref{fig:DCDM_Pk_e}, normalized to a $\Lambda$CDM reference model with all cosmological parameters fixed to Planck 2018 best-fit values. From top to bottom we increase the $\epsilon$ parameter, increasing the fraction of DR produced in the decay and decreasing the mass of the WDM daughter particle. The parameters $\epsilon$ and $\tau$ clearly have two different effects. For $\epsilon\ll 0.5$ the main effect of this parameter is to control the amount of rest mass that is transformed to kinetic energy in the decay, setting the free-streaming scale $k_\text{fs}$ of the WDM daughter particle. The lighter the particle as compared to its DCDM mother, the larger its kinetic energy, and the smaller is $k_\text{fs}$. This imprints a suppression in the power spectrum since structure is washed out for $k\gtrsim k_\text{fs}$. Thus, the position of the onset of suppression is determined by $\epsilon$ with lower values converging to CDM and therefore shifting the suppression more and more to the right. On the other hand, the decay time is responsible for the amount of WDM produced at a given time and thus controls the magnitude of the suppression.
	Note that the short oscillatory effect on large scales is caused by the switch to the fluid approximation and does not affect scales relevant for our analysis. These are marked by the pink region showing the $k$ ranges that are included in the BOSS Lyman-$\alpha$ data. 
	
	Depending on $\epsilon$ the suppression can start in three regimes. For low values it occurs on scales smaller than those probed by BOSS, implying that the power spectrum does not deviate strongly from $\Lambda$CDM (top panel in Fig.\,\ref{fig:DCDM_Pk_e}). On the contrary, for rather high $\epsilon$ values the suppression occurs already well before the BOSS region, such that the power spectrum is suppressed by an almost constant factor without scale dependence (bottom panel in Fig.\,\ref{fig:DCDM_Pk_e}). As we will discuss below, this shift can approximately be absorbed by astrophysical nuisance parameters within the effective model used here. Therefore, we do not expect a large difference to \lcdm either, as long as the suppression is not too extreme. The most interesting $\epsilon$ values are in the range between these limits for which BOSS can directly probe the shape of the power suppression and is therefore most sensitive (middle panel in Fig.\,\ref{fig:DCDM_Pk_e}). If there actually were a preference for DCDM by Lyman-$\alpha$ data we would expect an improvement of the fit over $\Lambda$CDM to occur in this parameter region.

	Finally, we stress that Lyman-$\alpha$ data are sensitive to the power spectrum at high redshifts (specifically $z=3.0-4.2$ for our analysis, see below). Due to the exponential decay law, this makes a significant difference in the amount of suppression in the power spectrum as compared to low redshifts. In Fig.\,\ref{fig:DCDM_Pk_zevolution} we show a DCDM spectrum for $\tau=20$\,Gyrs and $\epsilon=0.004$ at redshifts ranging from $z=0.1$ to $z=1000$, all normalized to the \lcdm spectrum at the respective redshifts. For large $z$, there is no difference because the decay has not set in yet, whereas at the lowest $z$ a large suppression occurs since the amount of WDM has increased and it had more time to wash out structures. The thick lines indicate the  relevant redshifts for our analysis at $z=3.0$ and $z=4.2$, which already show way less suppression of around $10-20\%$ compared to the $70\%$ for $z=0.1$. 
	
	As we are interested in the power spectrum on small scales we have to consider non-linear effects.
	While our treatment of the complex IGM physics entering the Lyman-$\alpha$ flux power spectrum is discussed in the next section, we first discuss our treatment of non-linearities related to three-dimensional matter clustering. Since we focus on the redshift range $z=3.0-4.2$, we can take advantage of the much milder non-linearities as compared to $z=0$, or equivalently the larger value of the non-linear scale. Indeed, three-dimensional matter clustering is still within the weakly non-linear regime for the redshifts and scales of the BOSS Lyman-$\alpha$ data. Following~\cite{Garny:2018byk, Garny:2020rom}, we therefore calculate the one-loop corrections from cosmological perturbation theory and use them as input in our analysis. As in~\cite{Simon:2022ftd}, we resort to the conventional EdS approximation for computing non-linear kernels, while it would be interesting to take the exact time dependence for DCDM into account following e.g. the strategy developed in~\cite{Garny:2020ilv}. In Fig.\,\ref{fig:DCDM_Pk_znl} we show the relative size of the one-loop corrections $P_{\delta\delta}/P_{\mathrm{linear}}$ for redshifts $z=0, 3.0$ and $4.2$ and $\tau=40$\,Gyrs, $\epsilon=0.006$ (solid) compared to \lcdm (dashed). We notice two effects: First, as expected, the corrections are much larger at $z=0$ since the non-linearities had more time to grow. In contrast, they are sufficiently small within the relevant range of scales and for the earlier redshifts we are interested in, justifying the perturbative treatment at the level of the matter power spectrum (see e.g.~\cite{Garny:2018byk} for a discussion of the impact of two-loop corrections). Second, for DCDM the corrections are overall smaller compared to \lcdm within the weakly non-linear regime because of the general suppression of growth, and the quadratic dependence on the linear input spectrum at one-loop order.
	
	\section{Lyman-\boldmath\texorpdfstring{$\alpha$}{alpha} forest data and model}
	\label{sec:Lya}
	
	\subsection{Data}
	\label{sec:data}
	
	To infer matter distributions in our universe the Lyman-$\alpha$ forest is a powerful tool for cosmological measurements in a relatively high redshift and small scale regime~\cite{Gnedin:1997td,Rauch:1998xn}.
	Measurements of the Lyman-$\alpha$ forest are useful in two regards. First, they can help studying the complex photo-ionized hot intergalactic medium. Second, the one-dimensional flux power spectrum is a powerful tool to constrain the underlying matter power spectrum on comparably small scales. High-resolution data, like those measured by HIRES (High Resolution Echelle Spectograph) from the KECK observatory or by MIKE~\cite{Viel:2013fqw,Vogt:1994fao} are sensitive to small fluctuations and provide a measurement up to very large wavenumbers $0.008-0.08$ s/km (around $1- 10h/$Mpc). These scales fall in the strongly non-linear regime and are largely affected by Jeans suppression from the IGM pressure. A modelling based on hydrodynamical simulations is indispensable for exploiting these data sets. Mid-resolution data like those provided by SDSS/BOSS~\cite{Palanque-Delabrouille:2013gaa,Chabanier:2018rga} are sensitive to larger scales, i.e. smaller wavenumbers $k\sim 0.001 - 0.02$ s/km (around $0.1- 2h/$Mpc), but compensate for the lower resolution by a much larger number of quasar spectra and correspondingly smaller error bars on the power spectrum. In addition, for the measured redshifts $z=2.2-4.6$ these scales correspond to the weakly non-linear regime, and are separated by around an order of magnitude from the baryonic Jeans scale. Therefore, BOSS data are potentially amenable to an effective field theory description for which the impact of the complex IGM can be encapsulated in a number of astrophysical nuisance parameters. This is indeed the strategy followed in this work, based on an effective model~\cite{Garny:2018byk,Garny:2020rom} that has been validated with hydrodynamical simulations~\cite{Bolton:2016bfs} for $\Lambda$CDM and massive neutrino cosmologies as well as a set of IGM parameters, and shown to provide a valid description of BOSS Lyman-$\alpha$ data while being able to absorb IGM uncertainties (temperature, adiabatic index, reionization history) in the effective parameters. We review this model below.
	
\begin{figure}[t]
		\centering
		\includegraphics[height=0.3\textheight]{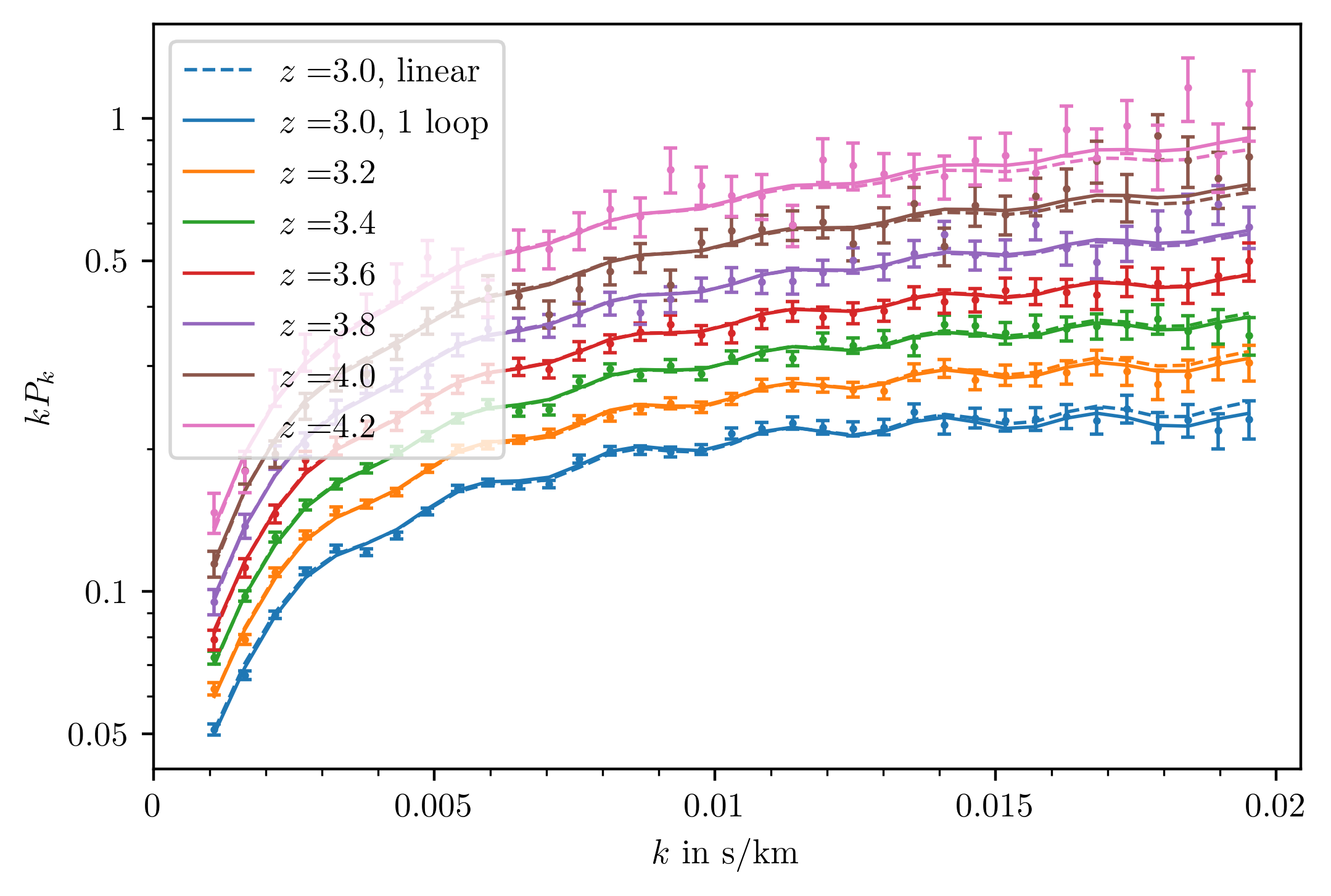}
		\caption{BOSS DR14 data for the one-dimensional Lyman-$\alpha$ forest flux power spectrum at redshifts $z=3.0-4.2$ used in this work, with errorbars as well as the linear (dashed) and one-loop (solid) best-fit effective model in the \lcdm case.}
		\label{fig:LCDM_Fit}
	\end{figure}

	We use the Lyman-$\alpha$ forest flux power spectrum from \cite{Chabanier:2018rga} based on BOSS DR14.
	We restrict ourselves to redshifts $z=3.0-4.2$ following the approach in~\cite{Garny:2020rom}, since lower redshifts are more sensitive to non-linearities and higher redshifts to reionization~\cite{Boyarsky:2008mt,2019MNRAS.489.3456G,Garzilli:2019qki}.
	We show the BOSS data in Fig.\,\ref{fig:LCDM_Fit} where we already included the effective model result for $\Lambda$CDM in the linear (dashed) and one-loop (solid) case to give an idea for the later result. The non-linear case clearly fits the data better which is also reflected in the lower $\chi^2$ value with $\Delta\chi^2=-13.4$.
	
	\subsection{Effective model}
	\label{sec:lyamodel}
	
	To use Lyman-$\alpha$ data for cosmological analyses we need to connect the one-dimensional flux power spectrum to the underlying three-dimensional matter power spectrum. This is often accomplished by running a set of hydrodynamical simulations corresponding to different cosmological and IGM parameters, and then constructing an interpolation among them~\cite{Palanque-Delabrouille:2015pga,Murgia:2018now,Hooper:2022byl,Pedersen:2022anu,Villasenor:2022aiy}. A simulation based strategy is indispensable for small-scale Lyman-$\alpha$ data. However, as mentioned above, for larger scale BOSS data it is also possible to obtain a valid description with a suitable effective model. We use the model discussed in~\cite{Garny:2020rom,Garny:2018byk}, that is based on a perturbative description combined with effective parameters that encapsulate the impact of the IGM, designed to work in the BOSS regime far above the baryonic Jeans scale and within the weakly non-linear regime of the underlying matter distribution. As described already above, the model has been validated with a set of hydrodynamical simulations~\cite{Bolton:2016bfs} and used to extract neutrino mass bounds as well as constraints on self-interacting dark matter models while marginalizing over IGM uncertainties. We review the setup here.
	
	The Lyman-$\alpha$ photon flux is determined by the transmission $F$ which depends on the optical depth $\tau$ via  $F=e^{-\tau}$. In particular we are interested in the fluctuations in the transmission spectrum
	\begin{equation}
		\delta_F = \frac{F}{\overline{F}} - 1,
	\end{equation}
	where $\overline{F}$ is the average transmission. 
	Since on BOSS scales the hydrogen clouds usually do not have large pressure gradients compared to the gravitational forces, and since damped Lyman-$\alpha$ systems are removed from the analysis, the underlying matter over- or under-densities can be traced by $\delta_F$. Therefore the optical depth and hence the transmission depend largely on the matter fluctuations $\delta$. Additionally, we have to account for the peculiar velocity $v_p$ and its gradient along the line of sight, generating distortions in the redshifts of the measured flux power spectrum. This can be described by a dependence on the dimensionless quantity
	\begin{equation}
		\eta = -\frac{1}{aH} \frac{\partial v_p}{\partial x_p}\,,
	\end{equation}
	with velocity $v_p$ and comoving coordinate $x_p$ along the line of sight.
	Now, we can expand the optical depth fluctuations at first order as~\cite{Gnedin:1997td} $\delta_\tau = b_{\tau\delta}\delta + b_{\tau\eta}\eta$, where we introduced the bias parameters $b_{\tau\delta}$ and $b_{\tau\eta}$. Applied to the transmission, we get in turn $\delta_F = b_{F\delta}\delta + b_{F\eta}\eta$ with the new parameters $b_{Fi} = \log(\overline{F}) b_{\tau i}$. 
	In linear approximation the velocity gradient would be proportional to the density contrast via $\eta = f\mu^2\delta$, where the $\mu^2=k_\parallel^2/k^2$ factor arises from only taking the contribution along the line of sight $k_\parallel$, and $f=d\ln D /d\ln a$ is the growth rate, with linear growth factor $D(a)$. In this case, we can compute the three-dimensional flux power spectrum $P_F(k) (2\pi)^3  \delta^3(\vec{k}+\vec{k'}) = \langle \delta_F(\vec{k}), \delta_F(\vec{k'})\rangle$ as
	\begin{equation}
		P_F^\text{lin}(k,\mu,z)  =  b_{F\delta}^2(z) \left(1 + \beta(z) \mu^2\right)^2  P_\text{lin}(k,z) \,,
		\label{eq:P_F linear}
	\end{equation}
	where we introduced the parameter $\beta(z) = f(z) b_{F\eta}(z)/b_{F\delta}(z)$, and $P_\text{lin}(k,z)$ is the  linear matter power spectrum. In the following, we omit the redshift arguments for brevity.
	
	To go beyond the linear approximation, we note that $\eta$ actually depends on the divergence of the three-dimensional velocity field $\vec{v}$ given by $\theta = \nabla\vec{v}/(a Hf)$ via $\eta = f\mu^2\theta$. Using this relationship we arrive at the generalized expression
	\begin{equation}
		P_F(k,\mu) = b_{F\delta}^2 \left( P_{\delta\delta}(k) + 2\beta\mu^2 P_{\delta\theta}(k) + \beta^2\mu^4 P_{\theta\theta}(k) \right)\,,
		\label{eq:P_F non-linear}
	\end{equation}
	with the additional dependence on the density and velocity power spectra $P_{\delta\delta}$ and $P_{\theta\theta}$ as well as the cross-correlation $P_{\delta\theta}$.
	The parameter $\beta$ can be estimated with the Zel'dovich approximation~\cite{1970A&A.....5...84Z}, where it only depends on the adiabatic index $\gamma$ that in turn is related to the reionization history~\cite{Hui:1997dp}. Since we do not want to impose any prejudice regarding the IGM, we model it with two free parameters $\alpha_\text{bias}$ and $\beta_\text{bias}$ allowing a power-law redshift dependence given by 
	\begin{equation}
		\beta(z) = \alpha_\text{bias}\left(\frac{a(z_\text{pivot})}{a(z)}\right)^{\beta_\text{bias}}\,,    
	\end{equation}
	choosing $z_\text{pivot} = 3.0$.
	
	In addition, there are also a number of other physical effects we need to account for. First of all, the evolution of baryonic matter is not only determined by gravitational forces but is also tied to its innate pressure. Unlike for dark matter, a collapse cannot happen below the Jeans scale $k_J = aH/c_s$, which is related to the sound velocity $c_s = \gamma T/(\mu_p m_p)$ with temperature $T$, adiabatic index $\gamma$ and the mean particle mass $\mu_p m_p$ of the IGM. More precisely, we have to look at the filtering scale $k_F$ which is given by a time average of the Jeans scale $k_J$~\cite{Gnedin:1997td},
	\begin{equation}
		\frac{1}{k_F^2(t)} = \frac{1}{D(t)} \int_{0}^{t} \mathrm{d}t' \frac{a^2(t')}{k^2_J(t')}
		\left[
		\frac{\mathrm{d}}{\mathrm{d}t'} \left( a^2(t') \frac{\mathrm{d}}{\mathrm{d}t'}D(t') \right)
		\right] \int_{t'}^{t} \frac{\mathrm{d}t''}{a^2(t'')}.
	\end{equation}
	For larger $k>k_F$, a suppression is modeled by an exponential factor $\exp(-\left( k/k_F\right) ^2)$. A typical value is of order $15-20 h/$Mpc, about an order of magnitude larger than the largest wavenumbers probed by BOSS. Therefore, Jeans suppression has only a minor impact on the relevant scales.

	Secondly, the spectral lines are subject to thermal broadening, leading to a damping of the flux power spectrum along the line of sight. This damping is also enhanced by other effects like redshift space distortions due to velocity dispersion as well as the finite resolution of the measurements~\cite{Garny:2018byk,Scoccimarro:2004tg}. To account for this, we include an overall exponential suppression factor of $\exp(-\left( k_\parallel/k_s\right)^2 )$ with the suppression scale $k_s$ being mainly determined by the thermal broadening, so $k_s \approx \sqrt{m_p/T}$. For example, for a temperature for the IGM of $T\approx 10^4$K, this yields $k_s \approx 0.11 $ s/km, corresponding to around $10 h/$Mpc. Thus, the effect of broadening is also subdominant within the BOSS regime.
	
	The last additional effect we need to account for are absorption features imprinted by other transitions than Lyman-$\alpha$ on the measured spectrum.
	The dominant effect stems from $\mathrm{SiIII}$ absorption, that can be accounted for by an oscillatory factor with wavelength $\Delta V = 2 \pi/0.0028$ km/s due to interference effects~\cite{Palanque-Delabrouille:2013gaa}. These effects are well constrained by the BOSS measurements, and can be described by
	\begin{equation}\label{eq:kappa}
		\kappa_{\mathrm{SiIII}} = 1 + 2\left( \frac{f_{\mathrm{SiIII}}}{1-\overline{F}} \right) \cos(\Delta V k_\parallel) + \left( \frac{f_{\mathrm{SiIII}}}{1-\overline{F}} \right)^2,
	\end{equation}
	where we also include the oscillation strength $f_{\mathrm{Si}III}=6\cdot10^{-3}$. The underlying transmission function entering $\kappa_{\mathrm{SiIII}}$ has only a very minor impact and we therefore use a fixed value $\log(\bar{F})(z)=-0.0025 (1 + z)^{3.7}$ within~\eqref{eq:kappa}~\cite{Garny:2020rom,Palanque-Delabrouille:2013gaa}. We stress that $\bar{F}$ is fixed \emph{only} within the factor described by~\eqref{eq:kappa}. This description was found to be sufficient and allowing the parameters $\Delta V$ and $f_{\mathrm{Si}III}$ to vary would not lead to relevant differences in the result since they are very well determined by the oscillatory features in the BOSS data~\cite{Garny:2020rom,Garny:2018byk}.
	
	To finally arrive at the one-dimensional spectrum, we integrate along the two directions that are not along the line of sight $k_\parallel$ leading to
	\begin{equation}
		P_{F,1D}(k_\parallel, z) = \frac{1}{2\pi} \int_{k_\parallel}^{\infty} k\mathrm{d}k P_F(k,\mu,z)\,,
	\end{equation}
	where $\mu=k_\parallel/k$.
	The integration can be performed for each of the power spectra in~\eqref{eq:P_F non-linear} which leads in turn to the three integrals
	\begin{flalign}
		I_0(k_\parallel,z) &= \int_{k_\parallel} \mathrm{d}k k \exp\left[-\left( \frac{k}{k_F}\right) ^2\right] P_{\delta\delta}(k,z), \\
		I_2(k_\parallel,z) &= \int_{k_\parallel} \mathrm{d}k \frac{k_\parallel^2}{k} \exp\left[-\left( \frac{k}{k_F}\right) ^2\right] P_{\delta\theta}(k,z)\,, \\
		I_4(k_\parallel,z) &= \int_{k_\parallel} \mathrm{d}k \frac{k_\parallel^4}{k^3} \exp\left[-\left( \frac{k}{k_F}\right) ^2\right] P_{\theta\theta}(k,z).
	\end{flalign}
	We observe that the powers of $k$ in the integrals are changing with the powers of $\mu\propto 1/k$ and we also have already included the Jeans suppression that also depends on $k$. As one can see, the integrals are in principle uncapped and potentially sensitive to extremely non-linear scales. In practice, for $I_2$ and $I_4$ these scales are not giving a relevant contribution since the integrand is strongly suppressed for large $k$ due to $(i)$ the inverse powers of $k$ and $(ii)$ the relatively mild enhancement or even suppression of the cross and velocity power spectra relative to the linear spectrum. For $I_0$ however, a relevant contribution from UV-scales arises. Note that for all $k_\parallel$ values relevant for BOSS the impact of the UV contribution can be absorbed by an additive constant~\cite{Garny:2018byk,Garny:2020rom}. To account for the UV contribution, we therefore include an extra additive counterterm $I_{ct}$ for $I_0$, which absorbs the uncertainty from the integration over these UV scales.  Again, we allow for a redshift dependence with 
	\begin{equation}
		I_{ct} = \alpha_\text{ct} \left( \frac{a(z)}{a(z_\text{pivot})}\right)^{\beta_\text{ct}} \,,
	\end{equation}
	and add thus two additional free parameters.
	As mentioned above, we use the $\delta\delta$, $\delta\theta$ and $\theta\theta$ auto or cross-correlation spectra computed at one-loop order in perturbation theory. 
	
	Lastly, we add an overall amplitude $A$ accounting for the overall bias factor $b_{F\delta}^2$ in~\eqref{eq:P_F non-linear} that we parameterize by
	\begin{equation}
		A(z)=\alpha_F \left( \frac{a(z_{pivot})}{a(z)}\right)^{\beta_{F}}\,,
	\end{equation}
	and include the thermal broadening. 
	Finally, with all the additional factors and the integrals,
	we arrive at
	\begin{flalign}
		P_{F,1D}(k_\parallel,z) = &A(z) \kappa_{\mathrm{SiIII}}(k_\parallel,z)  \exp\left[-\left(\frac{k_\parallel}{k_s(z)}\right)^2\right] \nonumber\\
		&\left(I_0(k_\parallel,z) + I_{ct}(z) + 2\beta(z) I_2(k_\parallel,z) + \beta(z)^2 I_4(k_\parallel,z)\right).
	\end{flalign}
	In practice, we use overall six free parameters
	\begin{equation}\label{eq:freeparam}
		\left\lbrace \alpha_F, \beta_F, \alpha_\text{bias}, \beta_\text{bias}, \alpha_\text{ct}, \beta_\text{ct} \right\rbrace\,,
	\end{equation}
	to account for the impact of the IGM and absorb uncertainties from UV modes when integrating across the line of sight.
	The IGM properties are determined by mainly the velocity bias parameters (shortened now to $\alpha_{b}$ and $\beta_{b}$) as well as $\alpha_F$ and $\beta_F$, while the non-linearities  are captured by the counterterm parameters $\alpha_\text{ct}, \beta_\text{ct}$. As discussed above the impact of the precise value of $k_s$ and $k_F$ is minor, and can moreover be compensated to a large extent by the other free parameters of the model~\cite{Garny:2020rom}. We therefore use fixed values $k_s=0.11$ s/km and $k_F=18 h/$Mpc (for a check of the dependence on this choice see~\cite{Garny:2020rom} and below).
	
	The advantage of this model is that it is very agnostic regarding the complex IGM physics which  leads to robust results when marginalizing over the free model parameters. This implies that the model leads per design to conservative constraints on cosmological parameters.

	\subsection{Fitting procedure}
	\label{sec:procedure}
	
	\begin{table}[t]
		\centering
		\begin{tabular}{|c|c|}
			\hline
			$\omega_b$ & 0.02237 \\
			\hline
			$\omega_\text{cdm}$ & 0.1200 \\
			\hline
			100$\Theta_s$ & 1.04110 \\
			\hline
			$\log(10^{10}A_s)$ & 3.044 \\
			\hline
			$n_s$ & 0.9649 \\
			\hline
			$\tau_\text{reio}$ & 0.0544 \\
			\hline
			\hline 
			$\sigma_8$ \ (\text{for}\ $\Lambda$CDM) & 0.810 \\
			\hline
			$S_8$ \ (\text{for}\ $\Lambda$CDM) & 0.833 \\
			\hline
		\end{tabular}
		\caption{Cosmological parameters for the \lcdm and DCDM model used in our analysis, fixed to Planck values~\cite{Planck:2018vyg}. In case of DCDM, $\omega_\text{cdm}$ is replaced with $\omega^\text{ini}_\text{dcdm}$. In addition, we use one massive neutrino species with $m=0.06$\,eV and two massless species. The $\sigma_8$ and $S_8$ values are model dependent and correspond to \lcdm. 
		}
		\label{tab:LCDM_Planckdata}
	\end{table}
	
	To determine the compatibility of a given model with the BOSS DR14 Lyman-$\alpha$ data set~\cite{Chabanier:2018rga} we follow a frequentist approach in this work, based on the profile likelihood. We first compute the linear power spectrum, then the one-loop density, velocity and cross power spectra as discussed above, and next the integrals $I_{0,2,4}$ for each of the 35 $k$ values and the seven redshift bins $z=3.0,3.2,\dots,4.2$. For each of these $245$ data points we evaluate the difference between the theoretical model and the measured value, and compute a $\chi^2$ taking the statistical uncertainties reported by~\cite{Chabanier:2018rga} into account. We then minimize $\chi^2$ with respect to the Lyman-$\alpha$ effective model parameters~\eqref{eq:freeparam}. To mitigate the impact of local minima the minimization is performed multiple times scanning different parameter range combinations. Additionally, we checked the robustness with respect to the model assumptions and repeated the fits with different cutoff scales. This not only tests the cutoff independence (by absorbing the cutoff dependence of $I_0$ into the values of the counterterm parameters) but provides also an extra check for possible minimization errors. In addition, we performed checks on the dependence on the fixed values for $k_s$ and $k_F$ as described in section \ref{sec:lyamodel} and found the result to only deviate below 1\%.
	
	We compare our DCDM results to a \lcdm reference model with the parameters specified in Tab.\,\ref{tab:LCDM_Planckdata} and results shown in Fig.\,\ref{fig:LCDM_Fit}. Additionally, we include one massive neutrino species with $m=0.06$\,eV and two massless species. This yields a baseline value of $\chi^2=192.89$ in the one-loop \lcdm fit. As noted before~\cite{Garny:2020rom}, the absolute $\chi^2$ value should be regarded with care, and we only use the relative $\chi^2$ difference for model comparison. 
	
	In order to explore the impact of Lyman-$\alpha$ data, for DCDM we fix the cosmological parameters to the same values as in the \lcdm reference model, postponing a full analysis to future work. While this choice certainly represents a limitation of our analysis that should be kept in mind, we observe that the corresponding results when taking CMB, BAO and SN Ia data into account do not show any strong degeneracies of the DCDM parameters $\tau,\epsilon$ with the other cosmological parameters~\cite{FrancoAbellan:2021sxk,Simon:2022ftd}. The main reason is that DCDM behaves identical to \lcdm around recombination, and when fixing the angular diameter distance to the last scattering surface (ensured by using $\Theta_s$ as input parameter). The parameter $\omega_\text{cdm}$ is replaced by the equivalent $\omega^\text{ini}_\text{cdm}$ such that the CDM densities also agree around recombination, before the decay sets in, as do the baryon,  photon and neutrino densities. It is therefore reasonable to expect that the cosmological parameters take the same values preferred by Planck in DCDM and \lcdm, respectively.
	To span a decent amount of parameter space, we choose 23 different lifetimes $\tau$ as well as 28 different $\epsilon$ values.\footnote{we use a grid with $\tau =  1, 2, 3, 4, 5, 6, 8, 10, 12, 15, 17, 19, 20, 23, 27, 30, 40, 50, 80, 100, 200, 500, 1000 \,\mathrm{Gyrs}$ and \\
		$\epsilon =  0.0001, 0.00015, 0.0002, 0.0003, 0.0004, 0.00055, 0.0007, 0.001, 0.0015, 0.002, 0.003, 0.004, 0.006, 0.008, 0.01, \\0.015, 0.02, 0.03, 0.04, 0.05, 0.065, 0.08, 0.1, 0.15, 0.2, 0.3, 0.4, 0.5$.
	} This results in overall 644 different parameter combinations.

	\section{Results}
	\label{sec:results}
	\subsection{Exclusion bounds}
	\label{sec:exclusionbounds}
	
	\begin{figure}[t]
		\centering
		\includegraphics[width=1\textwidth]{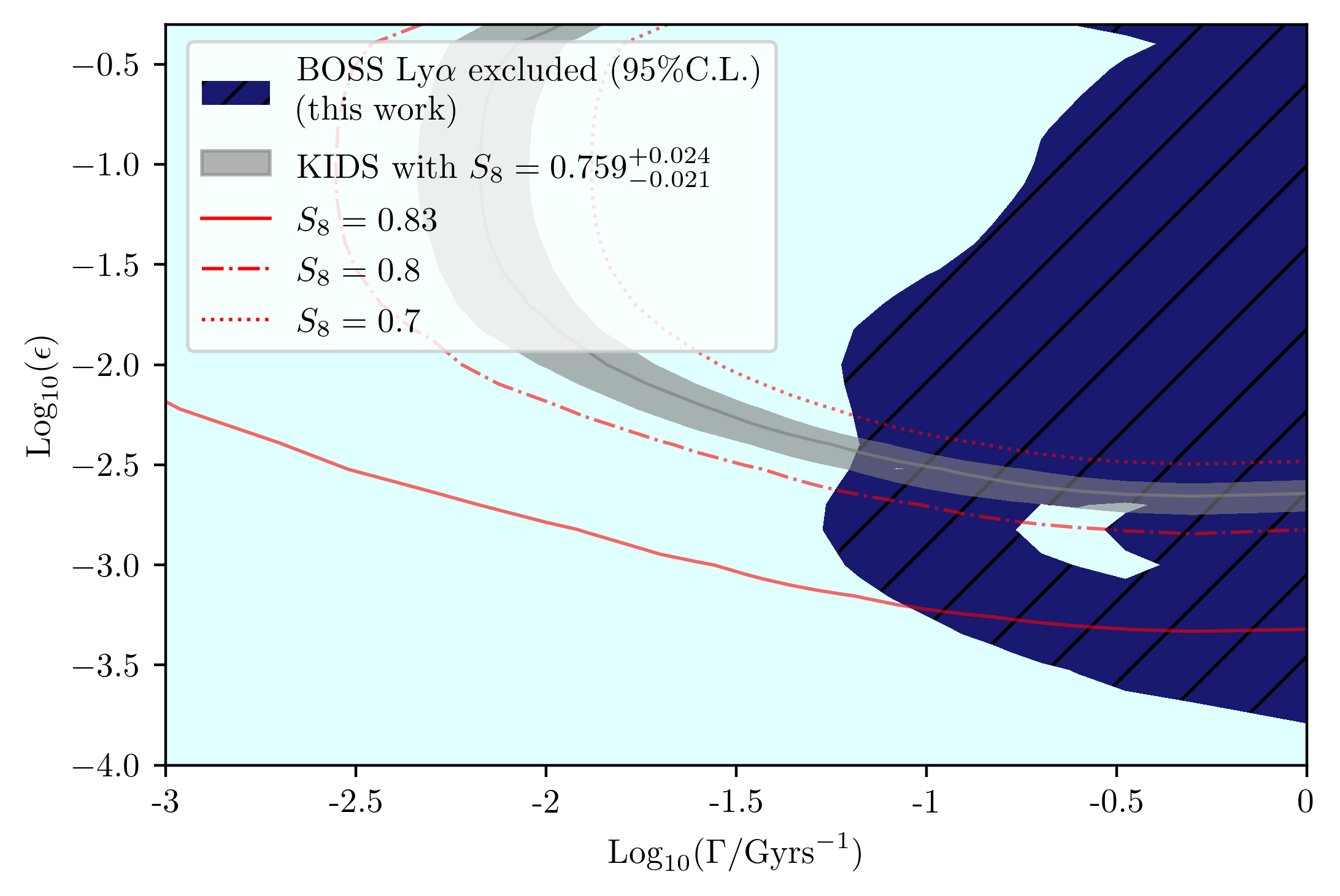}
		\caption{Excluded parameter space by BOSS DR14 Lyman-$\alpha$ data at $95\%$C.L. (dark blue and hatched). The red lines indicate contours of fixed $S_8$ within the DCDM model  
			and the gray band shows the region consistent with the value $S_8 = 0.759_{-0.021}^{+0.024}$ favoured by KIDS~\cite{KiDS:2020suj}.}
		\label{fig:DCDM_exclusion_ext}
	\end{figure}

	We find that DCDM is not statistically preferred over \lcdm by BOSS DR14 Lyman-$\alpha$ data even though the fit partly improves over \lcdm with a best-fit value of $\chi^2=189.77$ and $\Delta\chi^2=-3.1$ for $\tau\simeq 40$\,Gyrs and $\epsilon\simeq 0.006$. The region in the two-dimensional parameter space spanned by the decay rate $\Gamma=\tau^{-1}$ and the degeneracy parameter $\epsilon$ that is excluded at $95\%$C.L. when compared to the \lcdm reference model (i.e. has $\Delta\chi^2> 3.841$) is shown in Fig.\,\ref{fig:DCDM_exclusion_ext}.
	At the bottom left corner, DCDM converges to \lcdm since it corresponds to large lifetime and low $\epsilon$, meaning colder DM. Similarly, DCDM approaches \lcdm for both $\tau\to\infty$ and $\epsilon\to 0$, i.e. models close to the bottom and left axis are most \lcdm-like. This is also observed in the fit with a $\chi^2$ value close to the one of \lcdm when moving in those regions.
	
	At the right side of Fig.\,\ref{fig:DCDM_exclusion_ext}, the lifetime is very short, down to $1$Gyr. The suppression in the power spectrum is therefore very strong in this regime which leads to it being excluded for all $\epsilon\gtrsim 10^{-4}$. For even smaller $\epsilon$ the power suppression only arises for scales below those tested by BOSS Lyman-$\alpha$ data. 
	
	The bound on the lifetime is strongest for values of $\epsilon\sim 10^{-2}-10^{-3}$, where the $k$-dependent suppression of the power spectrum due to DM decay falls within the BOSS range. We find that a lifetime of $\tau \lesssim 18$\,Gyrs is approximately excluded at $95\%$C.L. in this region. For large values $\epsilon\gtrsim 0.1$, corresponding to large mass splitting, our analysis yields again weaker bounds on the lifetime since the power suppression is approximately $k$-independent within the BOSS window for these scales, such that it can be compensated by a corresponding shift in the opacity bias parameter (or equivalently $\alpha_F$ within our model parameterization). One could improve the Lyman-$\alpha$ bound in this case by imposing a prior on $\alpha_F$, for example due to a calibration of the opacity bias with hydrodynamical simulations, similar as was done for the neutrino mass analysis in~\cite{Garny:2020rom}. However, as we see below, this large $\epsilon$ region is already tightly constrained by CMB and BAO data, and therefore we stick to a conservative setup without imposing strong priors on the model parameters.
	One peculiar feature is the patch on the lower right where one small island in the parameter space is allowed. This is due to a small fluctuation of the $\chi^2$ value such that it is barely not excluded.
	
	Apart from the excluded region, we also show some contour lines of constant $S_8$ in Fig.\,\ref{fig:DCDM_exclusion_ext}, assuming Planck 2018 cosmological parameters as given in Tab.\,\ref{tab:bestfit}. Larger values are generated at the lower left corner and lower values at the upper right, due to an increasing level of suppression from DM decay. In addition, in the upper left corner the large mass splitting and low lifetime correspond to the regime where the change in background evolution for DCDM becomes relevant. This leads to a partial compensation of the power suppression due to a decreased Hubble rate at intermediate redshifts (see Fig.\,\ref{fig:DCDM_bg}), that leads in total to a relative enhancement of the growth rate. This explains the shape of the $S_8$ contours.
	The grey band shows the region in parameter space where the value of $S_8$ within DCDM is compatible with the KiDS range $S_8 = 0.759_{-0.021}^{+0.024}$ \cite{KiDS:2020suj}. Thus, we find that Lyman-$\alpha$ exclusion bounds do allow for low $S_8$ values within the DCDM model.

	\subsection{Allowed region and comparison with CMB and BAO data}
	\label{sec:comparison}
	
	\begin{figure}[t]
		\centering
		\includegraphics[width=1\textwidth]{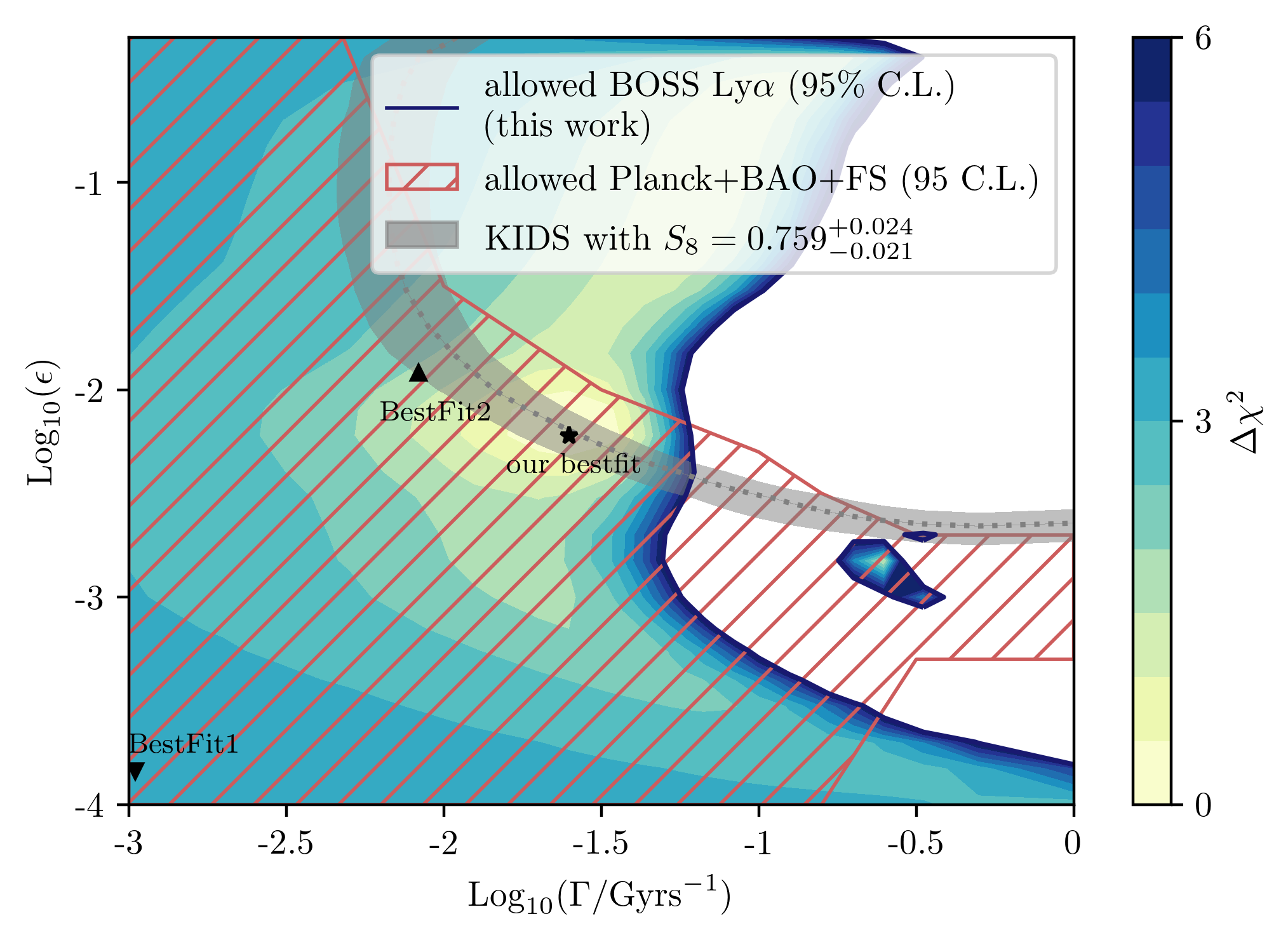}
		\caption{Allowed region at $95\%$C.L. resulting from Ly-$\alpha$ forest BOSS DR14 data (shaded, coloured according to $\Delta\chi^2=\chi^2-\chi^2_\text{min}$), around the Ly-$\alpha$ best-fit value (marked by a star). The hatched region is allowed at $95\%$C.L. by Planck, BAO, BOSS DR12 galaxy clustering full-shape (FS), and Pantheon data (referred to as Planck+BAO+FS), adapted from~\cite{Simon:2022ftd}. The gray band shows the KiDS range $S_8 = 0.759_{-0.021}^{+0.024}$~\cite{KiDS:2020suj}. BestFit1 and BestFit2 correspond to the best-fit points for Planck+BAO+FS and Planck+BAO+FS+KiDS reported in~\cite{Simon:2022ftd}, respectively, see Tab.\,\ref{tab:bestfit}.}
		\label{fig:DCDM_bestfit_contour}
	\end{figure}
	
	\begin{table}[t]
		\centering
		\begin{tabular}{|c||c||c|}
			\hline
			& BestFit1 & BestFit2  \\
			& Planck+BAO+FS & +KiDS $S_8$ \\
			\hline
			\hline
			$\log_{10}(\Gamma)$ & -2.98 & -2.08 \\
			\hline
			$\log_{10}(\epsilon)$ & -3.84 & -1.92 \\
			\hline
			\hline
			100$\omega_b$ & 2.245 & 2.242 \\
			\hline
			$\omega_\text{dcdm}^\text{ini}$ & 0.1190 & 0.1192 \\
			\hline
			$H_0$ & 67.82 & 67.73 \\
			\hline
			$\log(10^{10}A_s)$ & 3.051 & 3.052 \\
			\hline
			$n_s$ & 0.9679 & 0.9670 \\
			\hline
			$\tau_\text{reio}$ & 0.0584 & 0.0584 \\
			\hline
			\hline
			$\Omega_m$ & 0.3089 & 0.3094 \\
			\hline
			$\sigma_8$ & 0.823 & 0.763 \\
			\hline
			\hline
			$\chi^2$ & 3927.0 & 3929.3 \\
			\hline
			\hline
			$\chi^2$ from Ly-$\alpha$ & 193.26 & 191.76 \\
			\hline
			$Q_\text{DMAP}$  & \multicolumn{2}{c|}{$\sqrt{\chi^2_\text{w KiDS} - \chi^2_\text{w/o KiDS}}$ = 0.9$\sigma$} \\
			\hline
		\end{tabular}
		\caption{Parameter values and $\chi^2$ for the best-fit points from Planck+BAO+FS (BestFit1) and Planck+BAO+FS+KiDS (BestFit2) reported in~\cite{Simon:2022ftd}, as well as the additional contribution to $\chi^2$ from BOSS DR14 Lyman-$\alpha$ data analysed in this work and the $Q_\text{DMAP}$ statistic quantifying the compatibility with KiDS, when including also Lyman-$\alpha$ data. Due to the proximity of BestFit2 with the Lyman-$\alpha$ best-fit point, the tension is slightly lowered when including the results of this work (compared to $Q_\text{DMAP}=1.5\sigma$ for DCDM without Lyman-$\alpha$ data). For comparison, $Q_\text{DMAP}\simeq 3\sigma$ for \lcdm~\cite{Simon:2022ftd}.}
		\label{tab:bestfit}
	\end{table}
	
	In Fig.\,\ref{fig:DCDM_bestfit_contour} we show the region in DCDM parameter space that is allowed at $95\%$C.L. by BOSS DR14 Lyman-$\alpha$ data, corresponding to the region with $\Delta\chi^2=\chi^2-\chi^2_\text{min}\leq 5.991$ around the best-fit point at $\tau\simeq 40$\,Gyrs and $\epsilon\simeq 0.006$ (marked by a star). We note that for each point in parameter space we minimize $\chi^2$ with respect to the astrophysical parameters of the effective Lyman-$\alpha$ model, i.e. compute the profile likelihood to obtain the allowed region. For comparison we also show the Bayesian posterior obtained from an analysis of Planck, BAO, Pantheon and full shape BOSS DR12 galaxy clustering data (referred to as Planck+BAO+FS) reported in \cite{Simon:2022ftd}. Although the comparison should be treated with care due to the different statistical method and treatment of remaining cosmological parameters, it is instructive to see that the allowed regions are partially complementary to each other. In particular, there exists a region in parameter space for $\epsilon\sim 10^{-3}$ that is allowed by Planck+BAO+FS, but not allowed by Lyman-$\alpha$ data. Within this regime, the suppression of the power spectrum occurs on scales probed by the BOSS Lyman-$\alpha$ measurements.
	
	In addition, we show in Fig.\,\ref{fig:DCDM_bestfit_contour} the region in parameter space favoured by the value of $S_8$ measured by KiDS as gray shaded region. We see that there is an overlap of Lyman-$\alpha$, KiDS and Planck+BAO+FS allowed regions. Furthermore, it is intriguing that the KiDS band overlaps with the best-fit point of the Lyman-$\alpha$ analysis. Moreover, this point (marked with a star) is also close to the best-fit point from Planck+BAO+FS+KiDS data identified in~\cite{Simon:2022ftd}, shown by the upward pointing triangle in Fig.\,\ref{fig:DCDM_bestfit_contour}. This indicates that Lyman-$\alpha$ data are well compatible with the DCDM scenario that is preferred for relaxing the $S_8$ tension, and even slightly favours it.
	
	To further quantify the ability of DCDM to alleviate the $S_8$ tension, we consider the two best-fit points obtained from Planck+BAO+FS as well as Planck+BAO+FS+KiDS data~\cite{Simon:2022ftd}, referred to at as BestFit1 and BestFit2. In Tab.\,\ref{tab:bestfit} we show the corresponding model parameters as well as the $\chi^2$ values taken from~\cite{Simon:2022ftd}. In addition, we compute the contribution to $\chi^2$ from the BOSS DR14 Lyman-$\alpha$ analysis performed in this work. The slight preference of BestFit2 leads to a small reduction of the $Q_\text{DMAP}$ statistic quantifying the $S_8$ tension within DCDM from $1.5\sigma$ to $0.9\sigma$ when taking also Lyman-$\alpha$ data into account. For comparison, within \lcdm the $S_8$ tension quantified in this way is $\sim 3\sigma$.
	
	In summary, we find that BOSS DR14 Lyman-$\alpha$ data allow for values of the DM lifetime and mass splitting of mother and daughter particle that are favourable for resolving the $S_8$ tension between CMB and LSS data. This can be attributed to the strong redshift dependence of power suppression arising from DM decay, such that low $S_8$ values at $z\lesssim 1$ can be compatible with Lyman-$\alpha$ measurements at $z\sim 3-4$.
	
	\section{Three-body decay}
	\label{sec:3bodydecay}

	So far, we have studied a two-body decay with one massive and one massless daughter particle. However,  realistic dark matter models may allow only for decays with more particles in the final state due to selection rules imposed by underlying (approximate) internal and space-time symmetries. This frequently occurs for unstable particles in the Standard Model, such as for example for muons and neutrons, or more generally in nuclear $\beta$ decays. Therefore, we explore the changes when considering three- instead of two-body decays within the dark sector in this section (see e.g.~\cite{Blackadder:2014wpa} for some discussion in this direction). Specifically, we consider the decay
	\begin{equation}
		\text{DCDM} \ \rightarrow \ \text{WDM} \ + \ \text{DR}_a\ + \ \text{DR}_b\,,
	\end{equation}
	featuring one massive daughter particle (WDM) with $E_1=\sqrt{m^2+p_1^2}$  and now \emph{two} massless particles (DR$_a$ and DR$_b$) with $E_2=p_2$ and $E_3=p_3$, respectively. Whereas the momentum and energy in the two-body decay are fixed as described in~\eqref{eq:pmax}, an additional daughter particle leads to a continous momentum and energy distribution. 
	In the following we outline the changes in the formalism to account for this situation, and propose a simplified mapping that allows one to approximately translate results for the two-body decay to more general scenarios in the most relevant limit $\epsilon\ll 0.5$ (i.e. $m/M\to 1$). However, let us start by discussing the general three-body setup.
	
	The total decay rate for a three-body decay is given by
	\begin{equation}
		\Gamma = \frac{1}{(2 \pi)^3}\frac{1}{2}\frac{1}{8 M} \int \dx E_1 \dx E_3 \overline{|\mathcal{M}^2|}\,,
		\label{eq:gammatot}
	\end{equation}
	where one can choose freely over which two out of the three energies one integrates. Here $\overline{|\mathcal{M}^2|}$ is the matrix element squared, averaged (summed) over initial (final) state degrees of freedom. For concreteness, and in order to be agnostic about the precise origin of the three-body decay, we consider the spectrum that results purely from three-body kinematics, assuming that the matrix element squared for the decay can be approximated by a constant.
	
	The decay spectra are given by the differential cross sections $\dx \Gamma/\dx E_i$ for $i=1,2,3$, respectively. To compute it for the WDM particle ($i=1$), we use the minimal and maximal energies $E_3$ for a given $E_1$ allowed by energy and momentum conservation,
	\begin{align}
		\frac{\dx \Gamma}{\dx E_1} = {\cal N} \int \frac{\dx E_3}{M} &= \frac{\cal N}{M} \left( E_3(E_1)|_\text{max} - E_3(E_1)|_\text{min}\right) \nonumber \\
		&= \frac{ {\cal N}}{2M} \left(M - E_1+\sqrt{E_1^2-m^2}\right) - \frac{ {\cal N}}{2M} \left(M-E_1-\sqrt{E_1^2-m^2}\right) \nonumber \\
		&=  {\cal N}\frac{\sqrt{E_1^2-m^2}}{M}\,,
		\label{eq:dGdE1}
	\end{align}
	where we introduced the normalization factor ${\cal N}\equiv \overline{|\mathcal{M}^2|}/(128\pi^3)$.
	
	For the massless daughters ($i=2,3$), the shape of the spectra as dictated by kinematics are identical. For $i=3$ it is given by 
	\begin{align}
		\frac{\dx \Gamma}{\dx E_3}= {\cal N}\int \frac{\dx E_1}{M} &= \frac{\cal N}{M} \left( E_1(E_3)|_\text{max} - E_1(E_3)|_\text{min} \right) \nonumber \\
		&= {\cal N}\left(\frac{m^2+M^2}{2 M^2} - \frac{M^2 + m^2 - 4 E_3 (M - E_3)}{2 M^2 - 4 M E_3} \right)\nonumber \\
		&= \frac{{\cal N}}{M} \left(E_3 + \frac{E_3 M^2}{2E_3 M - M^2}\right)\,,
		\label{eq:dGdE3}
	\end{align}
	and for $i=2$ one has $\dx\Gamma/\dx E_2=\dx\Gamma/\dx E_3|_{E_3\to E_2}$. The total decay width takes the form
	\begin{equation}
		\Gamma = \int \frac{\dx \Gamma}{\dx E_1}\dx E_1 = \int \frac{\dx \Gamma}{\dx E_3}\dx E_3 = \frac{\cal N}{4} M \left(2\epsilon (1-\epsilon) +\left(1-2\epsilon\right) \log (1-2 \epsilon)\right)\,.
		\label{eq:gammatoteps}
	\end{equation}
	We show decay spectra for the massive (WDM) and one of the massless (DR) daughter particles in Fig.\,\ref{fig:3body}.
	For convenience we show the distribution with respect to the comoving momentum (see below) instead of the energy.
	For the value $\epsilon=0.499$ close to the maximal value $0.5$ (bottom right), the massive daughter is almost massless and also behaves as DR. Thus, all distributions converge towards each other in that limit. For the lower value of $\epsilon=0.4$ (lower left), the difference between the distributions becomes visible, and for $\epsilon\ll 0.5$ (upper row) the DR spectrum is peaked at half of the maximal possible momentum, while the WDM distribution always has its maximum at the maximally possible momentum. This can be understood by kinematics and phase-space arguments: the DR$_a$ particle reaches its maximal energy (and thus maximal absolute momentum) when the WDM and the DR$_b$ particles have momentum vectors that are pointing opposite to DR$_a$. Energy and momentum conservation require the momentum $p_3$ of DR$_b$ to approach zero in that limit. However, the number of available final states in phase-space is suppressed by a factor $p_3^2$ for $p_3\to 0$, explaining the suppression of the DR spectrum close to the endpoint. For WDM a similar restriction does not exist, such that it has a spectrum that remains finite at the endpoint.
	
	\begin{figure}[t]
		\centering
		\includegraphics[height=0.225\textheight]{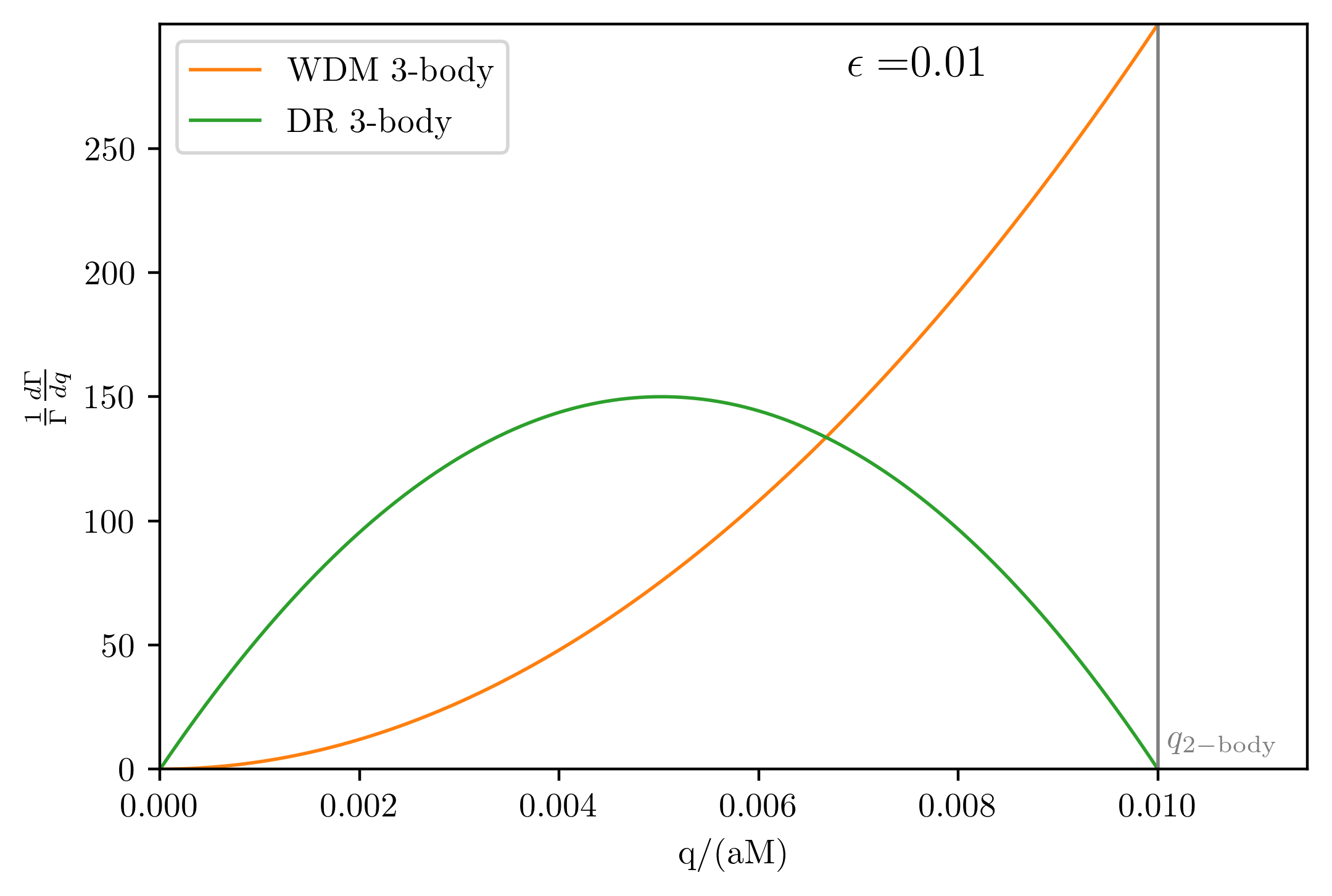}
		\\
		\includegraphics[height=0.23\textheight]{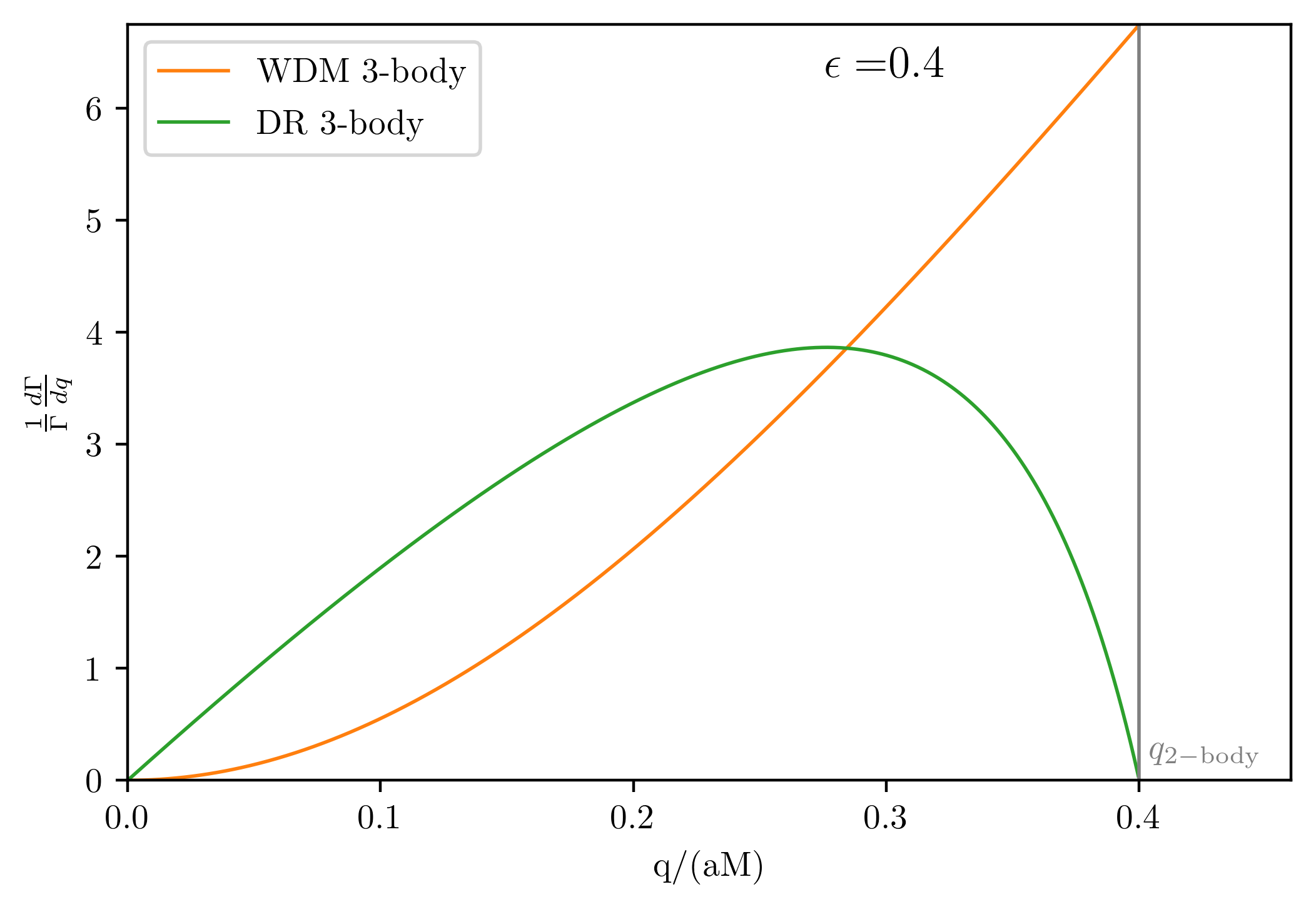}
		\includegraphics[height=0.23\textheight]{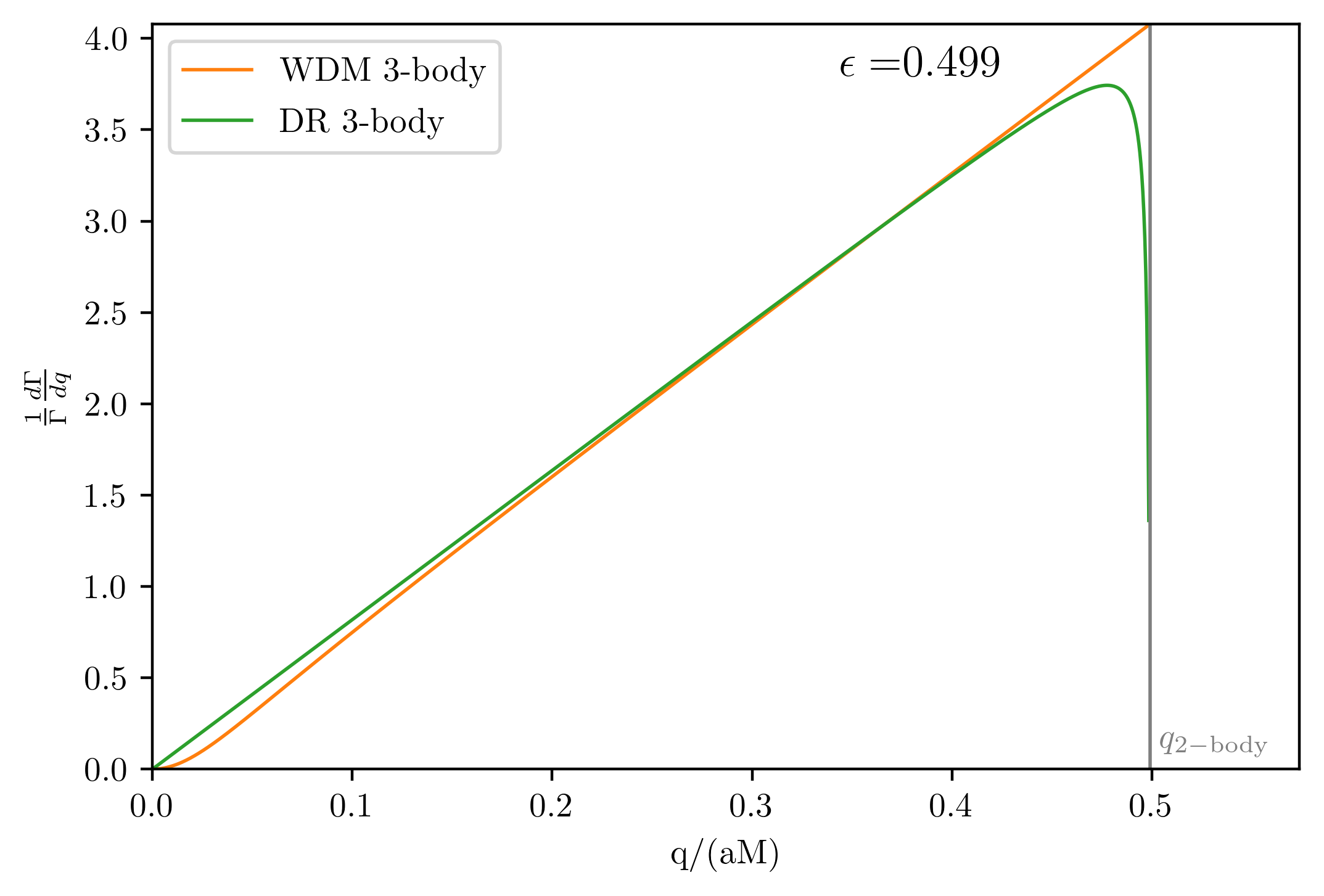}
		\caption{Momentum distribution of WDM and one DR particle in the three-body decay $\text{DCDM} \ \rightarrow \ \text{WDM} \ + \ \text{DR}_a\ + \ \text{DR}_b$ for the three different values $\epsilon=0.01, 0.4, 0.499$ of the mass splitting parameter, corresponding to $m/M=0.04, 0.45, 0.99$. The solid grey line shows the maximum momentum a particle in the three-body decay can have, being identical to the fixed momentum in the two-body case.
			For $\epsilon\to 0.5$, the distributions converge towards each other since all particles effectively behave as DR. For lower $\epsilon$ values the distributions differ due to phase-space suppression (see text for details).}
		\label{fig:3body}
	\end{figure}
	
	\begin{figure}[t]
		\centering
		\includegraphics[height=0.225\textheight]{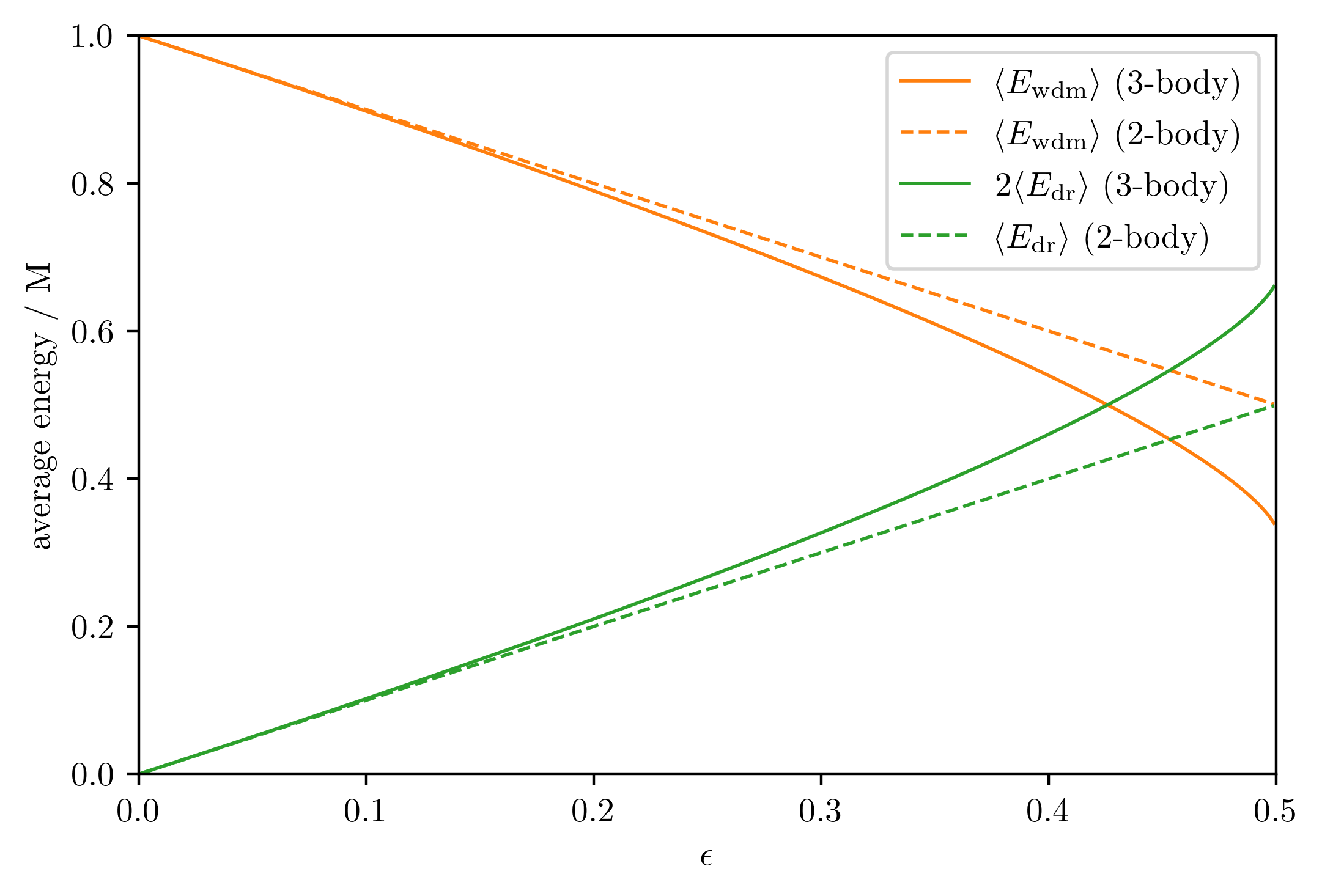}
		\includegraphics[height=0.225\textheight]{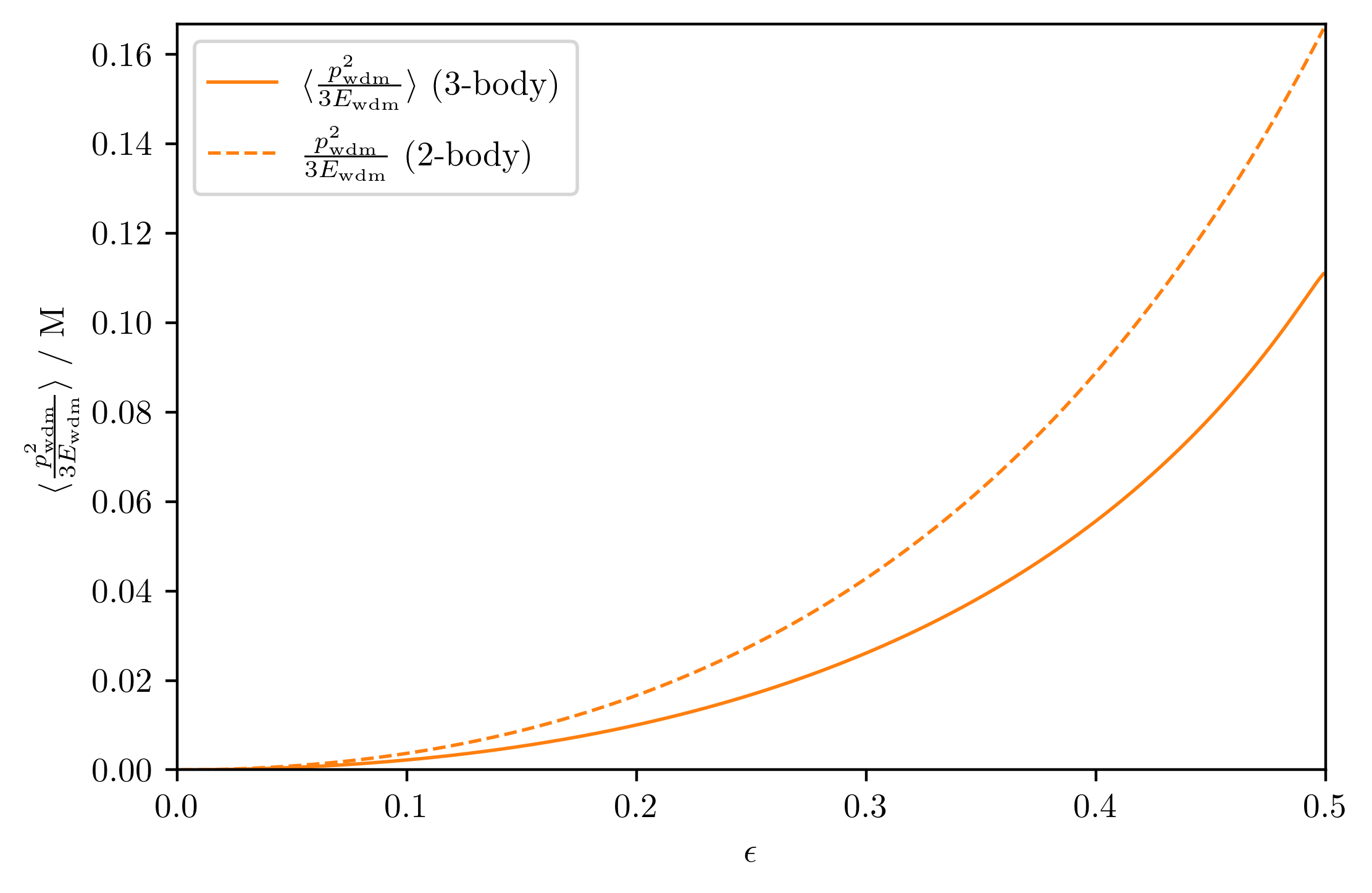}
		\caption{Left: Average energy $\langle E_\text{wdm}\rangle$ of the WDM (orange) and both DR particles $2\langle E_\text{dr}\rangle$ (green) in the three-body decay (solid lines), depending in the mass splitting parameter $\epsilon=(1-m^2/M^2)/2$. For comparison we also show the fixed energies for a two-body decay, given by $(1-\epsilon)M$ and $\epsilon M$, respectively (dashed). Right: same for the average of $\left\langle p_\text{wdm}^2/3E_\text{wdm}\right\rangle$ for three-body decay (solid) and the two-body case (dashed, given by $\epsilon^2/(3-3\epsilon)M$).}
		\label{fig:3bodyaverages}
	\end{figure}
	
	For later use we also define the average of some quantity $X$ (e.g. the energy of one of the decay products or some power of the momentum) by
	\begin{equation}
		\langle X\rangle \equiv \frac{1}{\Gamma}\int X d\Gamma
		= \frac{1}{\Gamma}\int dE_i X \frac{d\Gamma}{dE_i}\,.
	\end{equation}
	Note that in the last relation there is \emph{no} summation, but $i$ is fixed and it holds for any choice of $i=1,2,3$. Due to energy conservation one has $\langle E_1+ E_2+ E_3\rangle=M$, and since the massless daughters have identical spectra, one has $\langle E_2\rangle=\langle E_3\rangle$. We hence use the notation $\langle E_\text{wdm}\rangle\equiv\langle E_1\rangle$ and $\langle E_\text{dr}\rangle\equiv \langle E_2\rangle=\langle E_3\rangle$ for the average WDM and DR energies in the three-body decay. Energy conservation then implies $\langle E_\text{wdm}\rangle+2\langle E_\text{dr}\rangle=M$.
	Using the three-body spectrum from above, one finds
	\begin{equation}
		\frac{\langle E_\text{wdm}\rangle}{M} = 1-\frac{2\langle E_\text{dr}\rangle}{M} = \frac{4}{3}\frac{\epsilon^3}{2\epsilon (1-\epsilon) +\left(1-2\epsilon\right) \log (1-2 \epsilon)}\,.
		\label{eq:avenergy}
	\end{equation}
	We show the dependence of the average energies on $\epsilon$ by solid lines in the left panel of Fig.\,\ref{fig:3bodyaverages}. For $\epsilon\to0.5$ (i.e. $m\to 0$) one has $\langle E_\text{wdm}\rangle/M\to \langle E_\text{dr}\rangle/M\to 1/3$, consistent with the fact that all decay products become massless in that limit. For $\epsilon\to 0$ (i.e. $m\to M$) one has $\langle E_\text{wdm}\rangle/M\to 1-\epsilon$ and $\langle E_\text{dr}\rangle/M\to \epsilon/2$, i.e. as expected most of the energy goes into the rest mass of the massive daughter, with only little kinetic energy left over. In that limit the average energies agree with the fixed values of the energy in the two-body case (dashed lines in the left panel of Fig.\,\ref{fig:3bodyaverages}).
	
	For later use we also give the average of $p^2/3E$ over the decay spectrum for WDM (relevant for the pressure),
	\begin{equation}
		\left\langle\frac{p_\text{wdm}^2}{3E_\text{wdm}}\right\rangle = \frac{4M}{9}\frac{3\left( 1-2\epsilon\right)^{3/2} \arctan\left( \frac{\epsilon}{\sqrt{1-2\epsilon}} \right) - \epsilon (3-\epsilon(6+\epsilon)) }{2\epsilon (1-\epsilon) +\left(1-2\epsilon\right) \log (1-2 \epsilon)}\,.
		\label{eq:avpressure}
	\end{equation}
	For $\epsilon\to 0$ one has $\langle p_\text{wdm}^2/3E_\text{wdm}\rangle/M\to \epsilon^2/5$ in the non-relativistic limit, and for $\epsilon\to0.5$ it approaches the relativistic limit $\langle E_\text{wdm}\rangle/3$, which together with $\langle E_\text{wdm}\rangle\to M/3$ yields $\langle p_\text{wdm}^2/3E_\text{wdm}\rangle/M \to 1/9$. This average is also shown in the right panel of Fig.\,\ref{fig:3bodyaverages}. For the fixed values of the momentum in the two-body decay one obtains $\epsilon^2/(3-3\epsilon)M$. We note that $\langle p_\text{wdm}^2/3E_\text{wdm}\rangle$ is thus smaller by a factor $3/5$ compared to the corresponding quantity in the two-body case, for $\epsilon\ll 0.5$.
	
	Let us now discuss how the evolution equations are changed when considering three-body decays.
	The Boltzmann equations for the homogeneous part of the distribution functions read (compare to~\eqref{eq:BE} for the two-body decay)
	\begin{flalign}
		\dot{\bar{f}}\dcdm(q,\tau) &= -a\Gamma\bar{f}\dcdm(q,\tau)\,, \nonumber \\
		\dot{\bar{f}}\wdm(q_1,\tau)  &= \frac{a\Gamma\bar{N}\dcdm}{4\pi q_1^2} \left(\frac{1}{\Gamma}\frac{d\Gamma}{dq_1}\right)\,,\nonumber\\
		\dot{\bar{f}}_{\text{dr},a}(q_2,\tau) &=\frac{a\Gamma\bar{N}\dcdm}{4\pi q_2^2} \left(\frac{1}{\Gamma}\frac{d\Gamma}{dq_2}\right)\,,\nonumber\\
		\dot{\bar{f}}_{\text{dr},b}(q_3,\tau) &=\frac{a\Gamma\bar{N}\dcdm}{4\pi q_3^2} \left(\frac{1}{\Gamma}\frac{d\Gamma}{dq_3}\right)\,,
		\label{eq:BE3body}
	\end{flalign}
	where $d\Gamma/dq_i = (d\Gamma/dE_i \times dE_i/dp_i \times dp_i/dq_i)|_{p_i=q_i/a} = (d\Gamma/dE_i \times p_i/(aE_i) )|_{p_i=q_i/a}$. While the equation for DCDM is unchanged, the equations for WDM and the two DR components contain the momentum spectrum. In particular, the expression in the bracket replaces the delta function $\delta(q-ap_\text{2-body})$ that occurs for two-body decays by the respective decay spectra that are normalized such that $\int dq_i \Gamma^{-1}\,d\Gamma/d q_i=1$ for $i=1,2,3$ when integrating over the comoving momentum. They depend only on the three-body kinematics, while the dependence on the lifetime enters via the prefactor containing $\Gamma=\tau^{-1}$.
	Since the two massless daughter particles have identical decay spectra, it is sufficient to consider the total DR distribution given by
	\begin{equation}
		\bar f\dr(q,\tau) \equiv f_{\text{dr},a}(q,\tau)+f_{\text{dr},b}(q,\tau)\,.
	\end{equation}
	By multiplying the distribution functions with the energies and integrating over $q$ we obtain evolution equations for the energy densities, 
	\begin{flalign}
		\dot{\bar{\rho}}\dcdm &= -3\mathcal{H} \bar{\rho}\dcdm - a\Gamma\bar{\rho}\dcdm\,, \nonumber\\
		\dot{\bar{\rho}}\dr &= -4\mathcal{H} \bar{\rho}\dr + (2\langle E_\text{dr}\rangle/M) a\Gamma\bar{\rho}\dcdm\,, \nonumber\\
		\dot{\bar{\rho}}\wdm &= -3(1+\omega)\mathcal{H} \bar{\rho}\wdm + (\langle E_\text{wdm}\rangle/M) a\Gamma\bar{\rho}\dcdm\,.
		\label{eq:densityDE3body}
	\end{flalign}
	These equations are a generalization of~\eqref{eq:densityDE} to three-body decays, and contain the energy averaged over the decay spectrum. Using that $\langle E_\text{wdm}\rangle=M-2\langle E_\text{dr}\rangle$ they are formally identical to~\eqref{eq:densityDE} when replacing $\epsilon$ in these equations by $2\langle E_\text{dr}\rangle/M$ (see~\eqref{eq:avenergy}). This may suggest a mapping of the three- to the two-body scenario at the background level. However, this putative mapping would not be adequate at the level of perturbations, that are relevant for the power suppression, and thus the Lyman-$\alpha$ analysis, as we argue below. However, we note that when considering the most interesting limit $\epsilon\ll 0.5$ (i.e. $m\to M$), the precise value of $\epsilon$ actually becomes irrelevant as far as the background densities are concerned. The reason is that in this limit the DR energy density becomes negligibly small (with $\rho_\text{dr}\propto\epsilon$), while the WDM component becomes non-relativistic with $\omega\propto \epsilon^2\to 0$. Furthermore, the average energy entering the evolution equation~\eqref{eq:densityDE3body} for the WDM energy density approaches the limit $\langle E_\text{wdm}\rangle\to m\simeq M$, independent of $\epsilon$. Therefore, the time-evolution of $\bar\rho\wdm$ depends only on the lifetime $\tau$ but not on $\epsilon$ for $\epsilon\ll 0.5$. Altogether, the WDM evolution equations~\eqref{eq:densityDE3body} and~\eqref{eq:densityDE} for the two- and three-body cases become identical to each other for $\epsilon\ll 0.5$. Given that this is true also for the DCDM component, this implies that for the relevant background energy densities there is no difference between two- and three-body decays for $\epsilon\ll 0.5$.
	
	On the level of perturbations, the information on the decay spectrum enters in general via a momentum-dependent collision term in the Boltzmann equations for WDM and DR. In this work we do not attempt a full solution of this case, but rather propose a simple prescription to approximately map the two-body results to the three-body case in the most relevant limit $\epsilon\ll 0.5$. In particular, we note that in this limit the DR component becomes irrelevant while the WDM part can be well approximated by a fluid description, as discussed above. The suppression of the power spectrum is then encoded in the effective sound velocity following~\cite{FrancoAbellan:2021sxk}, where it was argued that the adiabatic sound velocity complemented by a small correction factor yields results in congruence with the full Boltzmann hierarchy for $\epsilon\ll 0.5$ in the two-body case. In the following we assume that the dominant effect of going from two- to three-body decays can be captured by the modification of the adiabatic sound velocity, leaving a more detailed analysis to future work. It is given by
	\begin{equation}
		\begin{aligned}
			c_g^2 \equiv \frac{\dot{\bar P}\wdm}{\dot{\bar\rho}\wdm} = \omega &\left(\left(5- \frac{\mathfrak{p}}{\bar{P}\wdm} \right) - a\Gamma \frac{\bar{\rho}\dcdm}{\bar{\rho}\wdm} \frac{1}{\mathcal{H} M\omega} \left\langle \frac{p\wdm^2}{3E\wdm} \right\rangle \right) \\
			\cdot &\left( 3(1+\omega) - a\Gamma \frac{\bar{\rho}\dcdm}{\bar{\rho}\wdm} \frac{1}{\mathcal{H} M} \left\langle E\wdm\right\rangle \right)^{-1}\,,
		\end{aligned}
	\end{equation}
	where $\mathfrak{p}$ is the pseudo-pressure~\cite{Lesgourgues:2011rh}, being a higher moment of the distribution function. This result is a generalization of the one given in~\cite{FrancoAbellan:2021sxk} to the three-body case, and contains the average over the decay spectrum of the quantity given in~\eqref{eq:avpressure}. In the non-relativistic limit $\epsilon\to 0$ the contribution from pseudo-pressure becomes suppressed by a relative factor $\epsilon^2$, as does $\omega$. Furthermore, inspecting the evolution equation for the pressure itself we find that for small $\epsilon$
	\begin{equation}
		\frac{c_g^2|_\text{3-body}}{c_g^2|_\text{2-body}} = \frac{\langle p_\text{wdm}^2/3E_\text{wdm}\rangle}{p_\text{2-body}^2/3E_\text{2-body}} = \frac{4(1-\epsilon)}{3 \epsilon^2}\frac{3\left( 1-2\epsilon\right)^{3/2} \arctan\left( \frac{\epsilon}{\sqrt{1-2\epsilon}} \right) - \epsilon (3-\epsilon(6+\epsilon)) }{2\epsilon (1-\epsilon) +\left(1-2\epsilon\right) \log (1-2 \epsilon)}\,.
	\end{equation}
	This ratio approaches $3/5$ for $\epsilon\to 0$. Following the arguments from above we can approximately account for this reduction in the sound velocity parameter by re-interpreting the results of our Lyman-$\alpha$ analysis obtained in the two-body case for some given set of parameters $\epsilon'$ and $\tau$ as constraints that apply also to the three-body decay for a mass spectrum with $\epsilon=(1-m^2/M^2)/2$ and lifetime $\tau$ with the matching
	\begin{equation}
		\frac{{\epsilon'}^2}{3-3\epsilon'} = \frac{c_g^2|_\text{3-body}}{c_g^2|_\text{2-body}}\frac{\epsilon^2}{3-3\epsilon}\,,
	\end{equation}
	which corresponds to
	\begin{equation}
		\epsilon'(\epsilon) = \sqrt{\frac{3}{5}}\epsilon + \frac{\sqrt{15}-3}{10} \epsilon^2 + \frac{23\sqrt{15}-70}{700} \epsilon^3 + \mathcal{O}(\epsilon^4)\,.
	\end{equation}
	For $\epsilon\ll 0.5$ one has $\epsilon'=\sqrt{3/5}\epsilon\simeq \epsilon/1.3$. 
	The results from Fig.\,\ref{fig:DCDM_exclusion_ext} and Fig.\,\ref{fig:DCDM_bestfit_contour} can thus be interpreted as constraints on the parameters ($\tau,\epsilon'(\epsilon)$) for case of three-body decays.
	This means for example that the best-fit scenario obtained from our analysis of the BOSS DR14 Lyman-$\alpha$ data, corresponding to $\epsilon'=0.006$, can be translated from the two- to the three-body case by $\epsilon=1.3\epsilon'=0.0078$. The best-fit thus occurs for $m/M=0.994$ for the two-body decays, and for almost the same mass splitting $m/M=0.992$ in the three-body decay scenario.
	
	In summary, in this section we propose a simple mapping of results obtained from the two-body decay of CDM into one massive and one massless species onto a three-body decay into one massive and two massless species, that is approximately valid in the limit $\epsilon\ll 0.5$ ($m\lesssim M$). After observing that the evolution of the relevant CDM and WDM background energy densities are independent of $\epsilon$ in that limit, we determine a mapping by comparing the evolution equations for the WDM pressure, that enters in the fluid description of the WDM density perturbations adopted in~\cite{FrancoAbellan:2021sxk}. The evolution of the pressure depends on the average of $p_\text{wdm}^2/3E_\text{wdm}$ over the decay spectrum, which differs for two- and three-body decays. Since the pressure largely determines the free-streaming scale and thereby the suppression of the power spectrum, the simplified mapping should give a good indication of  how Lyman-$\alpha$ constraints can be translated from two- to three-body decays. Along similar lines, a mapping to decays involving even more particles in the final state could be obtained. In addition, all steps can easily be generalized to the case when the matrix element of the decay is not constant, but replaced by the proper matrix element within a specific particle physics model.

	\section{Conclusion}
	\label{sec:conclusion}
	
	In this work, we derive constraints on decaying dark matter from the BOSS DR14 one-dimensional Lyman-$\alpha$ forest power spectrum. Specifically, we consider a decay of cold dark matter into a massless dark radiation and a massive warm dark matter component. This setup has been discussed as a solution of the $S_8$ tension in view of CMB and LSS measurements. The decay leads to a time-dependent suppression of the matter power spectrum, leading to lower values of $S_8$ in the late universe, as indicated by a variety of weak lensing and cluster number count measurements, while preserving the success of \lcdm in explaining the CMB anisotropies and galaxy clustering on the largest scales. 
	
	The Lyman-$\alpha$ forest provides an important constraint for any model leading to a suppression of the power spectrum on scales $k\gtrsim 0.1 h/$Mpc. The pecularity of DCDM is that the suppression builds up over time due to the gradual decay. Therefore DCDM can potentially be consistent with low $S_8$ at $z\lesssim 1$, as hinted at e.g. by weak lensing, as well as Lyman-$\alpha$ forest constraints at  $z\sim 2-4$. Broadly speaking, our results confirm this expectation, making DCDM a viable candidate for solving the $S_8$ tension.
	
	We use an effective model for the one-dimensional Lyman-$\alpha$ forest flux power spectrum, that is designed to work on scales of BOSS data, far above the baryonic Jeans scale and within the weakly non-linear regime of the underlying matter density field. The effective model contains in total six free parameters that account for uncertainties from the IGM as well as the impact of non-linearities, and has been validated with hydrodynamical simulations for various cosmological models in the past.
	
	We find that for certain values of the mass degeneracy parameter $\epsilon=(1-m^2/M^2)/2$, BOSS Lyman-$\alpha$ data yield constraints on the dark matter lifetime $\tau$ that are \emph{stronger} compared to a combination of Planck, BAO, SN Ia and galaxy clustering data. The lower bound reaches $\tau\gtrsim 18$\,Gyrs for $\epsilon\sim 0.1-0.5\%$. Interestingly, the lifetime $\tau\sim 10^2\,$\,Gyrs and $\epsilon\sim 1\%$ that is favoured for relaxing the $S_8$ tension is also marginally preferred by BOSS Lyman-$\alpha$ data, as compared to \lcdm. The $S_8$ tension according to KiDS data is around $3\sigma$ within \lcdm, and is reduced to $1.5\sigma$ for DCDM. When including also BOSS Lyman-$\alpha$ data it is slightly further reduced to $\sim 1\sigma$. While the hint for this mild preference is intriguing, the main conclusion is that Lyman-$\alpha$ data are compatible with dark matter decay being a possible explanation of the $S_8$ tension.
	
	Apart from the question whether the $S_8$ tension is due to systematic effects, it would be interesting to investigate realistic and well-motivated models of dark matter decay. As a first step in this direction, we provide a mapping from the two-body decay scenario to a more general three-body decay that is valid in the limit $\epsilon\ll 0.5$ (i.e. for $m \to M$), taking the different phase space into account. The mapping amounts to a rescaling of the $\epsilon$ parameter. This allows one to easily translate constraints derived for the two-body case to models with three-body decays. We find that the best-fit Lyman-$\alpha$ scenario corresponds to $m/M=0.994$ for two-body decays, and $m/M=0.992$ for the three-body case.
	
	\acknowledgments{%
		
		We thank Guillermo F. Abell\'an for support regarding a modified version of CLASS accounting for DCDM, and Alejandro Ibarra, Vivian Poulin and Henrique Rubira for useful discussions.
		This work was supported by the DFG Collaborative Research Institution Neutrinos and Dark Matter in Astro- and Particle Physics (\href{"https://www.sfb1258.de/"}{SFB 1258}).
	}

	
\providecommand{\href}[2]{#2}\begingroup\raggedright\endgroup

\end{document}